\documentclass[acmtog]{acmart}

\acmSubmissionID{385}

\setcopyright{acmcopyright}
\acmJournal{TOG}
\acmYear{2025}
\acmVolume{44}
\acmNumber{4}
\acmArticle{1}
\acmMonth{8}
\acmDOI{10.1145/3730842}

\usepackage{color}
\usepackage{xcolor}
\usepackage{graphicx}
\usepackage{booktabs}
\usepackage{hyperref}
\usepackage{amsmath, amsthm, amsfonts, amssymb}
\usepackage{mathrsfs}
\usepackage{textcomp}
\usepackage{epstopdf}
\usepackage{multirow}
\usepackage{wrapfig}
\usepackage{booktabs} %
\usepackage[linesnumbered,ruled,vlined]{algorithm2e}
\usepackage{seqsplit}
\usepackage{ragged2e}
\usepackage[normalem]{ulem}
\usepackage{cleveref}
\usepackage{bm}
\usepackage{caption}
\usepackage{subcaption}
\usepackage{ulem}
\usepackage{enumitem}

\usepackage{booktabs, multirow} %
\usepackage{soul}%
\usepackage{xcolor} 
\usepackage{overpic}

\definecolor{green}{rgb}{0, 0.5, 0}
\definecolor{orange}{rgb}{0.6, 0.3, 0.1}
\definecolor{red}{rgb}{1.0, 0.0, 0.0}
\definecolor{teal}{rgb}{0.0, 0.4, 0.4}
\definecolor{purple}{rgb}{0.65,0,0.65}
\definecolor{saffron}{rgb}{0.95,0.75,0.2}
\definecolor{turquoise}{rgb}{0.0,0.4,0.8}
\definecolor{brown}{rgb}{0.5, 0.16, 0.16}
\definecolor{brickred}{rgb}{.6, .2 .1}
\definecolor{coral}{rgb}{1,0.45,0.33}
\definecolor{newcolor}{rgb}{.8,.349,.1}

\let\maketitlesup\maketitle
\usepackage{xpatch}
\xpatchcmd{\maketitlesup}{\@mkteasers}{}{}{}
\xpatchcmd{\maketitlesup}{\@mkabstract}{}{}{}
\xpatchcmd{\maketitlesup}{\@copyrightpermission}{}{}{}
\usepackage{float}
\usepackage{stfloats}

\begin{document}
\title{HoLa: B-Rep Generation using a Holistic Latent Representation}

\author{Yilin Liu}
\email{whatsevenlyl@gmail.com}
\affiliation{%
	\department{College of Computer Science \& Software Engineering}
	\institution{Shenzhen University}
	\country{China}	
}
\author{Duoteng Xu}
\email{duotengx@gmail.com}
\affiliation{%
	\institution{Shenzhen University}
	\country{China}	
}
\author{Xingyao Yu}
\email{yuxingyao50@gmail.com}
\affiliation{%
	\institution{Shenzhen University}
	\country{China}	
}
\author{Xiang Xu}
\email{samxucmu@gmail.com}
\affiliation{%
	\institution{Simon Fraser University}
	\country{Canada}	
}
\author{Daniel Cohen-Or}
\email{cohenor@gmail.com}
\affiliation{%
	\institution{Tel Aviv University}
	\country{Israel}	
}	
\author{Hao Zhang}
\email{hao.r.zhang@gmail.com}
\affiliation{%
	\institution{Simon Fraser University}
	\country{Canada}	
}
\author{Hui Huang}
\email{hhzhiyan@gmail.com}
\authornote{Corresponding author: Hui Huang (hhzhiyan@gmail.com)}
\affiliation{%
	\department{College of Computer Science \& Software Engineering}
	\institution{Shenzhen University}
	\country{China}	
}

\renewcommand\shortauthors{Y. Liu, D. Xu, X. Yu, X. Xu, D. Cohen-Or, H. Zhang, and H. Huang}

\begin{abstract}

We introduce a novel representation for learning and generating Computer-Aided Design (CAD) models in the form of {\em boundary representations\/} (B-Reps). 
Our representation unifies the continuous geometric properties of B-Rep primitives in different orders (e.g., surfaces and curves) and their discrete topological relations in a {\em holistic latent\/} (HoLa) space. This is based on the simple observation that the topological connection between two surfaces is intrinsically tied to the geometry of their intersecting curve. 
Such a prior allows us to reformulate topology learning in B-Reps as a geometric reconstruction problem in Euclidean space.
Specifically, we eliminate the presence of curves, vertices, and all the topological connections in the latent space by learning to distinguish and derive curve geometries from a pair of surface primitives via a neural intersection network.
To this end, our holistic latent space is only defined on surfaces but encodes a full B-Rep model, including the geometry of surfaces, curves, vertices, and their topological relations.
Our compact and holistic latent space facilitates the design of a first diffusion-based generator to take on a large variety of inputs including point clouds, single/multi-view images, 2D sketches, and text prompts.
Our method significantly reduces ambiguities, redundancies, and incoherences among the generated B-Rep primitives, as well as training complexities inherent in prior multi-step B-Rep learning pipelines, while achieving greatly improved validity rate over current state of the art: 82\% vs. $\approx$50\%.
\end{abstract}




\begin{teaserfigure}
  \centering
  \includegraphics[width=\linewidth]{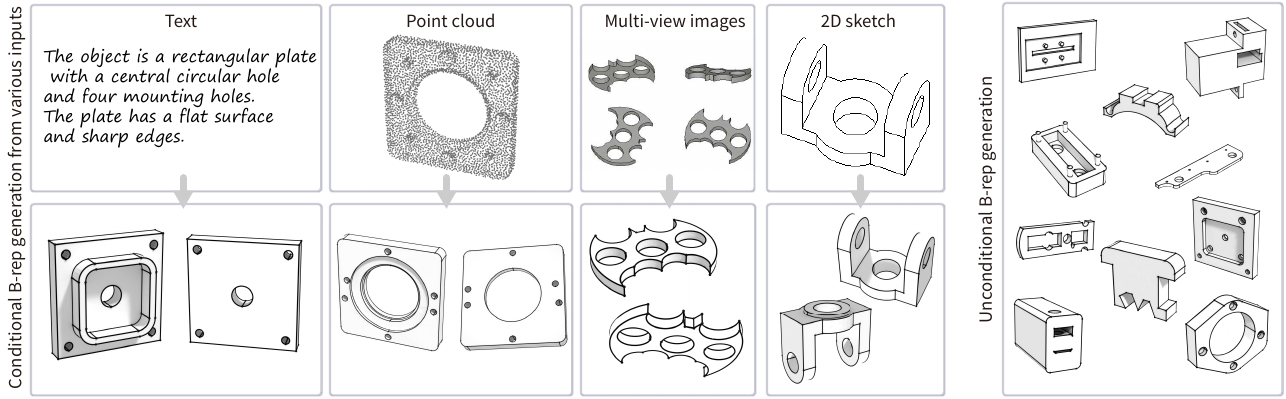}
  \caption{
  We learn a {\em holistic latent} (HoLa) space for generating boundary representations (B-Reps), which facilitates the design of our {\em first\/} diffusion-based generator to take on 
  a large variety of inputs. Conditional generation results of 3D B-Reps are shown, in two views (bottom), for text, point cloud, multi-view image, and 2D sketch inputs. Unconditional B-Reps generation
  with HoLa exhibits greatly improved validity over current state of the art: 84\% vs.~$\approx$50\%.
  }
  \label{fig:teaser}
\end{teaserfigure}

\maketitle

\section{Introduction}
\label{sec:intro}

3D content generation is a vital task in computer graphics and generative AI, gaining increasing importance in applications such as virtual reality, gaming, and manufacturing. 
While triangle meshes, point clouds, and more recently neural fields~\cite{NF_survey}, are widely used for 3D representation, researchers are turning to more compact and structured representations to facilitate editing, analysis, and manipulation. This is especially prevalent in domains such as design, engineering, robotics, and even eCommerce, which work with Computer-Aided Design (CAD) models. The Boundary Representation or B-Rep, consisting of parametric primitives (surfaces, curves, and vertices) and their topological relations, has been the most fundamental 3D shape format employed for CAD.
However, applying modern-day machine learning to B-Reps, in particular, for B-Rep generation, is challenging since the original B-Rep is characterized by a combination of {\em continuous\/} geometric parameters for its primitives and their topological relationships, which are {\em discrete\/} in nature. This necessitates the exploration of advanced representation learning approaches to B-Rep modeling and
generation.

\begin{figure}[!t]
  \centering
  \includegraphics[width=\linewidth]{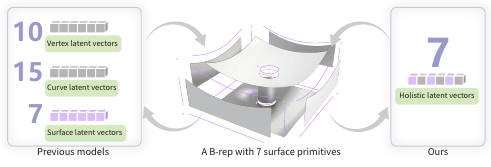}
  \caption{
        {\textbf{Disjoint vs.~holistic latent B-Reps.} Previous neural B-Rep models all learn {\em separate\/} latent spaces and their respective decoders/generators for surface, curve, vertex primitives, even their relations. In contrast, our HoLa representation is learned to blend the geometry and topology of all primitives into one unified latent space, for B-Rep generation.}
  }
  \label{fig:teaser2}
\end{figure}

Previous methods for B-Rep generation typically rely on multi-step training pipelines to address the discrete nature of the representation. 
A typical approach is to design {\em separate\/} decoder or generator networks for different primitive categories, i.e., surfaces, curves, and 
vertices, where the objective of each generator is {\em primitive formation\/} within a B-Rep, without explicitly enforcing any topological constraints between the 
generated primitives.
For instance, the curve generator in B-RepGen~\cite{brepgen24} seeks to generate curves that properly trim the input surface to contribute to a B-Rep, 
while the surface generator in SolidGen~\cite{solidgen23} needs to select a subset of input curves to form a B-Rep surface patch. 
These generators are both unaware of the key topological relation about whether two generated surfaces are connected through a generated curve.

The main issue with these generative models is the lack of explicit topology constraints or priors during training, together with error accumulation over their multi-step generation pipelines. 
As such, the generated B-Reps often contain ambiguities, incoherence, and redundancies among the resulting primitives. %
These would lead to invalid outputs or reliance on post-processing to fix the problems, e.g., de-duplication by B-RepGen to remove duplicated curves.

In this paper, we introduce a novel representation for learning and generating B-Reps. It unifies the {\em continuous\/} geometric properties of B-Rep primitives in different categories and their {\em discrete\/} topologies in a {\em holistic latent\/} (HoLa) space, as shown in Fig.~\ref{fig:teaser2}.
Our key observation is the simple fact that any curve in a B-Rep must be the intersection between two surface primitives.
As such, the topological connection between two higher-order primitives (i.e., surfaces) is intrinsically tied to the geometry of a lower-order primitive (i.e., a curve). 
This relationship provides a natural bridge to connect different primitive categories and their inner- and inter-topological relations within a unified representation;
it can well serve as a topo-geometrical prior or inductive bias when designing our autoencoder for learning a holistic latent, or HoLa.

Our latent space is defined over surface primitives. Specifically, given a pair of surface latents, we employ a \textit{neural intersection} module to identify their intersection and recover the geometry of the lower-order primitive, i.e., the intersection curve, during the decoding process.
By restricting topology learning to individual surface pairs and their corresponding intersecting curves, our neural architecture is significantly simplified (see Fig.~\ref{fig:overview}) compared
to those of prior models such as BRepGen and SolidGen.

We utilize a variational autoencoder (VAE) to learn the latent space for B-Rep generation. 
The encoder, comprising convolutional layers along with self- and cross-attention mechanisms, encodes the geometry of different primitives and their topological connections into our
holistic latent space for surfaces. 
Subsequently, intersection modules are employed to decode the geometry of lower-order (curve) primitives and their topological relationships. 
A reconstruction loss across all primitives and classification loss over all intersections supervise the training of our autoencoder network, ensuring consistent representations of both geometry and topology. 
Leveraging this unified latent space, we further train latent diffusion models (LDMs) to generate B-Rep models under various conditions, including noise (unconditional generation), single/multi-view images, sparse/dense/noisy point clouds, 2D sketches, and text prompts; see Fig.~\ref{fig:teaser} for a sampler of B-Rep generation results.

Extensive experiments and evaluations demonstrate that our method generates more plausible B-Rep models with a higher validity ratio compared to state-of-the-art alternatives, while also exhibiting greater diversity in the generated models. 
Unlike previous multi-step pipelines that require retraining separate models for surfaces, curves, and vertices, our unified latent space allows us to only retrain one diffusion model when input conditions change, making our approach significantly more efficient and effective.

\section{Related Work}
\label{sec:rw}

In this section, we review representation learning methods for B-Reps (Sec.~\ref{sec:rw-brep}) and their applications for CAD reconstruction and generation (Sec.~\ref{sec:rw-brepgen}). 
We also briefly discuss alternative CAD representations, such as Constructive Solid Geometry (CSG) and sketch-and-extrude techniques, in Sec.~\ref{sec:rw-others}. 
Finally, we highlight recent advancements in conditional CAD model generation in Sec.~\ref{sec:rw-condition}.

\subsection{Representation Learning of B-Reps}
\label{sec:rw-brep}

B-Rep is a crucial, and the {\em de facto\/}, format for CAD models~\cite{solidgen23}, directly encoding both their geometry and topology. 
A B-Rep comprises parametric surfaces, curves, points, and their topological connections, making it well-suited for graph-based representations~\cite{brep85}.
Over time, researchers have proposed using both \textit{homogeneous}~\cite{uvnet21,brepnet23,Cao20} and \textit{heterogeneous} graphs~\cite{AutoMate21,JoinABLe22,HGCAD23} to represent B-Rep models.

In a homogeneous graph, primitives such as surfaces are treated as nodes. For instance, UV-Net~\cite{uvnet21} and BRepNet~\cite{brepnet23} use graph convolutional or custom kernels to learn the attributes of B-Reps from surface nodes.
Smirnov et al.~\shortcite{template21} leverage a pre-defined topology template and deformable Coons patches~\cite{Coons67} to construct B-Reps.
Similarly, Jones et al.~\shortcite{ssr23} represent B-Reps using explicit surfaces and implicit curves in the parametric space to enable more continuous learning of discrete topological relationships.
However, homogeneous graphs are typically limited to tasks at the surface or body level, missing finer details such as edge loops and vertices. 

In contrast, heterogeneous graph representations, as employed by AutoMate~\cite{AutoMate21} and Bian et al.~\shortcite{HGCAD23}, encode different primitive types and their topological relationships both within and across categories, such as surfaces versus curves or surfaces versus parts. 
This approach is capable of capturing more advanced attributes, such as assembly relationships.
While heterogeneous graphs can effectively capture complex interrelationships between different classes of primitives, handling the diverse geometric types and topological connections also introduces significant challenges for downstream tasks such as 3D generation.

\subsection{3D reconstruction and generation using B-Rep}
\label{sec:rw-brepgen}

The lack of a unified representation for the diverse geometric and topological components in B-Rep models has driven the development of various approaches to address this challenge. 
Segmentation-based methods~\cite{parsenet20, primitivenet, hpnet21, bpnet23, sed23} cluster points based on learned features to segment the input point cloud into distinct surface primitives.
Non-segmentation methods, such as ComplexGen~\cite{complexgen22}, directly construct B-Rep models by detecting both primitives and their topological connections from point clouds. 
More recently, NVDNet~\cite{nvd24} introduces an intermediate representation, the Voronoi diagram, to encode both geometry and topology into a continuous form to improve generalizability.
While these approaches can recover the missing curves and vertices through post-processing methods such as Point2CAD~\cite{point2cad}, the decoupling of fine-grained geometric and topological details during training often leads to a lack of precision in the reconstructed models, limiting the quality and detail of the final B-Rep outputs.

The heterogeneous nature of B-Rep primitives also restricts the application of generative models to produce novel B-Rep shapes.
Previous works usually resort to a multi-step pipeline to circumvent this problem.
For example, SolidGen~\cite{solidgen23} iteratively generates vertices, curves, and surfaces, in that order\/, to construct a solid model, where a subsequent generation is conditioned on the output of the previous generation via pointer networks~\cite{pointer15}. The key problem related to topology learning is that of selecting \textit{several} subsets of lower-order primitives (e.g., curves) to form higher-order primitives (e.g., surfaces). This is considerably harder to train than our neural intersection module.

BRepGen~\cite{brepgen24} takes a surface latent to first generate a set of surfaces, and this step is completely topology unaware. A separate curve generator is employed, conditioned on each generated surface, to produce a set of B-Rep curves. Topology learning is conducted by forming {\em all\/} curves on each surface, with a topology tree employed to guide B-Rep generation. However, the topology of different primitives is implicitly encoded during the generation and is separated from the geometry learning process.
The lack of a unified representation for geometry and topology leads to much ambiguity and inconsistency in the generated B-Reps.

With respect to terminology, ComplexGen also adopted the term ``holistic'' to characterize their B-Rep chain complex representation that encompasses both B-Rep primitives and their connections.
However, their tri-path transformer decoder produces and operates over three {\em separate\/} groups of embeddings, one each for surfaces, curves, and vertices, where a
cross attention mechanism is developed to enforce coherence between the three types of primitives for 3D reconstruction. 
In contrast, our holistic representation embeds all the primitive geometry and topological properties into a single, unified latent vector, offering critical advantages to B-Rep generation.

\subsection{Constructive Solid Geometry and Sketch-and-Extrude}
\label{sec:rw-others}

Beyond B-Reps, researchers have explored Constructive Solid Geometry (CSG) as an alternative CAD representation. In CSG, shapes are constructed using basic primitives and pre-defined Boolean operations~\cite{Nandi17, Nandi18, DuIPSSRSM18, TianLSEFTW19, SharmaGLKM18, EllisNPSTS19, RitchieGJMSWW23, KaniaZK20, RenZ0LJCZPZZY21, bspnet20, capri22, d2csg23}. Neural CSG models are predominantly trained using reconstruction errors, but there are infinitely many CSG trees that would yield zero reconstruction error~\cite{nvd24}, hence there is no guarantee that the resulting CSG tree is optimal. In practice, the results from state-of-the-art neural reconstruction algorithms typically contain an  excessive number of redundant primitives.

Command-sequence-based methods, such as sketch-and-extrude techniques, are also widely used for 3D CAD reconstruction~\cite{you2024img2cad, LiG0Y23, XuPCWR21} and generation~\cite{RenZCLZ22, XuJLWF23, CADParser23, chen2024img2cad}. 
However, sequential CAD generative models are limited by the fixed order of the commands and the absence of explicit topological information, making it challenging to generate complex CAD models.
However, the geometric complexity of generated shapes is often limited by the error accumulation in the sequential generation and supported operations, particularly the constraints of the extrude operation. Moreover, accessing the construction history of generated models typically entails a more complex and costly data collection process. For instance, existing datasets for B-Rep models are approximately five times larger than those for sketch-and-extrude models, making large-scale training less feasible in certain scenarios.

\subsection{Conditional CAD model generation}
\label{sec:rw-condition}

Recent advances in generic 3D generation~\cite{clay} have significantly enhanced the ability to produce 3D models from diverse conditions, such as images, point clouds, and texts. However, most existing methods focus on triangle meshes and signed distance functions. While effective for many applications, these representations mostly fail to adequately capture  sharp features or complex topological structures inherent in CAD models.

In the domain of CAD generation, methods have primarily focused on generating or reconstructing CAD models from point clouds~\cite{parsenet20, complexgen22, nvd24, point2cad, UyCSGLBG22} or voxel grids~\cite{voxel2cad22}, which provide straightforward conditioning mechanisms for the generative processes. Conditioning on 2D inputs, such as single-view image~\cite{solidgen23, you2024img2cad, chen2024img2cad} or sketches~\cite{Sketch2CAD20, img2sketch}, introduces additional challenges due to the inherently ill-posed nature of recovering full 3D geometry from limited or ambiguous inputs.
Recent efforts, such as CAD-MLLM~\cite{xu2024cadmllm}, have introduced multimodal generative frameworks capable of generating CAD models from a variety of conditional inputs, further advancing the field. Nevertheless, a significant limitation of the majority of these methods is their reliance on command-based generation approaches, which can restrict their expressiveness and flexibility.
To our knowledge, no existing work directly generates B-Rep models from diverse conditional inputs as in our work.

\begin{figure*}
  \centering
  \includegraphics[width=\linewidth]{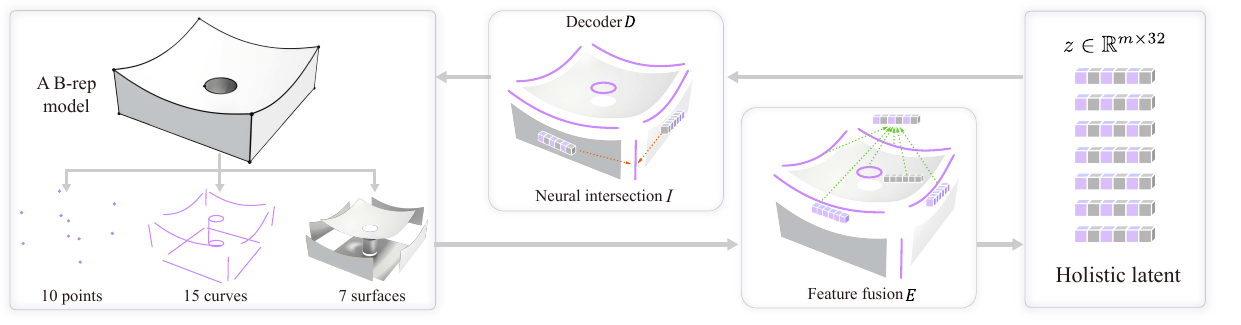}
  \caption{
  A B-Rep model that consists of surfaces, curves, and their topological connections is encoded into a holistic latent space via a feature fusion model.
  Then we emply a novel neural intersection module to recover the geometry surfaces and all the lower-order primitives from the latent space.
  }
  \label{fig:overview}
\end{figure*}

\section{Overview}
\label{sec:overview}

To generate B-Rep CAD models from various conditions, our framework consists of two main components: 1) a self-supervised variational autoencoder (VAE) that learns a holistic latent space for representing B-Rep models, and 2) a latent diffusion model (LDM) that generates latent vectors conditioned on various inputs (e.g., noise, point clouds, images, sketches or text descriptions) to produce B-Rep models, as shown in Fig.~\ref{fig:overview}.

\paragraph{Input B-Rep Model}
A B-Rep CAD model $(S,C,V,T_{SC},T_{CV})$ is represented as a set of parametric surfaces $\{S_i\}_{i=1}^{m}$, half-curves $\{C_j\}_{j=1}^{n}$ \footnote{in the following we use curve to denote half-curve for simplicity}, and vertices $\{V_k\}_{k=1}^{l}$, along with surface-to-curve connections $T_{SC}$ and curve-to-vertex connections $T_{CV}$. 
Following previous works~\cite{brepnet23,brepgen24}, each surface and curve is treated as a finite B-spline primitive and represented using uniform sample points in the UV-space ($S_i \in \mathbb{R}^{16 \times 16  \times 3}$ for surfaces and $C_i \in \mathbb{R}^{16 \times 3}$ for curves). 
Topological connections are encoded as binary adjacency matrices, with $\mathbf{T_{SC}} \in \{0,1\}^{m \times n}$ representing surface-to-curve connections and $\mathbf{T_{CV}} \in \{0,1\}^{n \times l}$ representing curve-to-vertex connections.

Although surface primitives represented by grid sample points include boundaries, these boundaries do not necessarily align with the ground-truth model. 
Curve primitives are required to trim surfaces and form watertight B-Rep models. 
Meanwhile, since the vertices can also be represented as the two endpoints of a curve, we merge vertices $\{V_k\}$ to curve $\{C_j\}$ to form the final representation $(S,C,T_{SC})$.

\paragraph{VAE Training}
We design a VAE to extract a holistic latent representation suitable for subsequent generation. 
The encoder network $\mathbb{E}$ processes the input B-Rep model $(S,C,T_{SC})$ by propagating vertex and curve features to surface primitives based on their topological connections, producing a latent vector $z_s \in \mathbb{R}^{m \times (2 \times 2 \times d)}$, where $z_s$ is defined per surface primitive with a spatial resolution of $2$ in the UV-space and feature dimension $d=8$. 
During decoding, the intersection module $\mathbb{I}$ recovers curve and vertex features from the sampled latent $z$, and the decoder network $\mathbb{D}$ reconstructs the input B-Rep model $(S,C,T_{SC})$ from the recovered features.

\paragraph{LDM Training}
Based on the learned latent representation, we train a latent diffusion model (LDM) $\mathbb{LDM}$ to generate B-Rep models conditioned on various inputs, such as single-view image, multi-view images, sparse or dense point clouds, single-view sketches and text descriptions. 
Conditions are processed by fixed or learned feature extractors to produce a $256$-dimensional condition vector $c \in \mathbb{R}^{256}$. 
When no condition is provided, the condition vector is set to $0$. 
The LDM then learns to denoise a fixed-length latent vector $\hat{z} \in \mathbb{R}^{M \times (2 \times 2 \times d)}$ from random Gaussian noise, conditioned on $c$, where $M$ is the predefined number of surface primitives in the B-Rep model.

\paragraph{Post-Processing}
The denoised latent vector $\hat{z}$ is fed into the intersection module $\mathbb{I}$ and decoder network $\mathbb{D}$ to generate the final B-Rep model $(S,C,T_{SC})$. 
Following methods like SolidGen~\cite{solidgen23} and BRepGen~\cite{brepgen24}, the output is converted into a watertight CAD model using OpenCascade~\cite{opencascade}.

\section{Methodology}
\label{sec:method}

\begin{figure*}
  \centering
  \includegraphics[width=\linewidth]{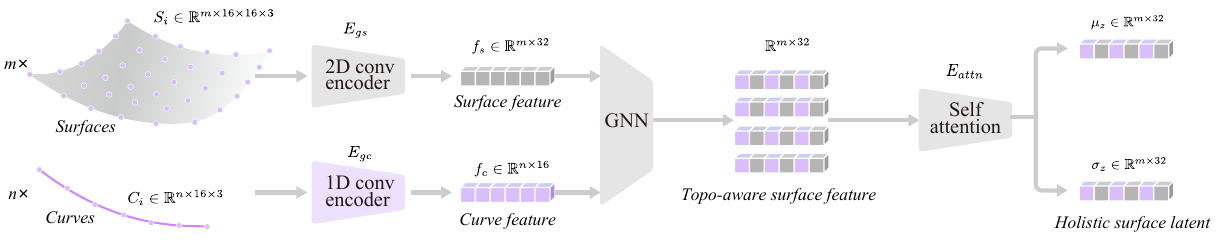}
  \caption{
 Given a GT B-Rep model, we use convolutional, self-attention and cross-attention layers to fuse the features of surfaces and curves and their connectivity to learn a holistic latent representation for the B-Rep model.
 }
  \label{fig:encoder}
\end{figure*}

\begin{figure*}
  \centering
  \includegraphics[width=\linewidth]{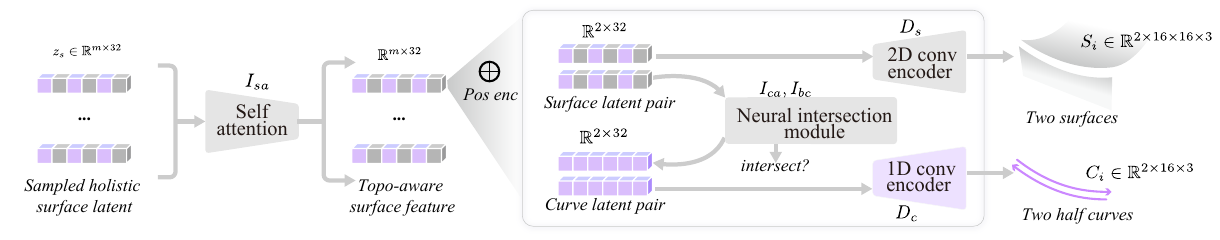}
  \caption{
 Given a pair of sampled surface latent $z_s$ from our holistic latent space, we use a neural intersection module to identify and recover the intersected curve feature. Then, the decoder network takes the sampled and recovered features to reconstruct the input B-Rep model $(S,C,T_{SC})$.
 }
  \label{fig:decoder}
\end{figure*}

While both the geometry and topology of B-Rep models are essential for learning a B-Rep model, previous methods only encoded the geometry in the latent space. They relied on the transition of generative models that specialized in different kinds of primitives to capture their topological connection implicitly.
However, unlike previous methods, we design a specific B-Rep variational autoencoder (VAE) to encode both the geometry and topology in a holistic latent space, enabling a more compact and expressive B-Rep representation.

The key to obtaining such latent space is: 1) encode both the continuous geometric over different primitive categories (surfaces, curves, and vertices) and discrete topological information to form a unified yet expressive latent vector, and more importantly, 2) decode the unified latent vector to recover not only surfaces but all types of primitives and their topological connections. 
We discuss those two steps in Sec~\ref{sec:method-encoder} and Sec~\ref{sec:method-decoder}.
Based on the learned latent, we show the design of the latent diffusion model (LDM) in Sec~\ref{sec:method-generation} to generate B-Rep both unconditionally and conditionally and the post-processing steps in Sec~\ref{sec:method-post} to convert the generated B-Rep into watertight CAD models.

\subsection{Encoder}
\label{sec:method-encoder}
To obtain a unified latent vector $z_s$ from a B-Rep model $(S,C,T_{SC})$, we design a VAE to encode the geometry of different primitives and their topological connections into a latent space for surfaces, as shown in Fig.~\ref{fig:encoder}.

\paragraph{Geometry Encoding:}
Given sample points of $m$ surfaces $\{S_i\}_{i=1}^{i=m} \in \mathbb{R}^{m \times 16 \times 16 \times 3}$ and $n$ curves $\{C_j\}_{j=1}^{j=n} \in \mathbb{R}^{n \times 16 \times 3}$, we first employ a set of convolution and downsample layers for surfaces $E_{gs}: \mathbb{R}^{m  \times 16 \times 16 \times 3} \rightarrow \mathbb{R}^{m \times 32}$ and for curves $E_{gc}: \mathbb{R}^{n \times 16 \times 3} \rightarrow \mathbb{R}^{n \times 16}$:
\begin{align}
 f_s = E_{gs}(S_i), 
 f_c = E_{gc}(C_i),
\end{align}
where $f_s \in \mathbb{R}^{m \times 32}$ and $f_c \in \mathbb{R}^{n \times 16}$ are the geometric feature vectors for surfaces and curves.
Note that the feature dimension of the surface is designed as $32=2 \times 2 \times 8$, where $8$ is the feature dimension and we preserve the \textit{spatial resolution} of $2$ to distinguish the direction of the primitive since the final max pooling operation is permutation invariant and will lose the orientation of the primitive.
This design has been used across all other feature vectors in our network.

\paragraph{Topology Encoding:}
Then we train a graph neural network~\cite{gatv2} $GNN: (\mathbb{R}^{m \times 32},\mathbb{R}^{n \times 16},\{0,1\}^{m \times n}) \rightarrow \mathbb{R}^{m \times 32}$ to propagate the curve features to the surface primitives based on the topological connection $T_{SC}$:
\begin{align}
 f_{cs} = GNN(f_s,f_c,T_{SC}),
\end{align}
where $f_{cs} \in \mathbb{R}^{m \times 32}$ is the curve-aware feature vector for the surface primitives.
We further reinforce the feature vector by employing a set of self-attention layers $E_{attn}: \mathbb{R}^{m \times 32} \rightarrow \mathbb{R}^{m \times 32}$ to aggregate and exchange features of individual surface latent in order to capture longer-term relation.
Finally, we map the topology-aware surface latent to the mean and variance of the latent Gaussian through an MLP layer: $E_{mlp}: \mathbb{R}^{m \times 32} \rightarrow (\mathbb{R}^{m \times 32}, \mathbb{R}^{m \times 32})$.
The final latent vector $z_s$ can be sampled from the predicted mean and variance.

\subsection{Decoder and Loss Function}
\label{sec:method-decoder}

A straightforward approach in previous works to supervise the training is to reconstruct the surface primitives $\{S_i\}_{i=1}^{m}$ and conduct $L1$ or $L2$ loss between the reconstructed and the input surface.
However, this will not guarantee that the latent vector will encode the topological information and all other primitives.
To enforce the latent vector to encode the holistic information about a B-Rep model, we design a neural intersection module $\mathbb{I}$ to recover the geometry of lower-order primitives (curves and points) and their topological relationships based on the surface latent $z_s$, as shown in Fig.~\ref{fig:decoder}.

\paragraph{Intersection Module:}
We observed that the intersection of two surfaces is always tied to their boundary curves.
We use this observation to inject the topological and curve information into the latent space.
Given a pair of sampled surface latent, we train a neural intersection module $I: (\mathbb{R}^{32},\mathbb{R}^{32}) \rightarrow (\mathbb{R}^{16}, \{0,1\})$ to recover the feature of the intersected curve and predict whether the two surfaces intersect or not.

Specifically, we first use a set of self-attention layers to further exchange the features of different surface latent after latent sampling $I_{sa}: \mathbb{R}^{m \times 32} \rightarrow \mathbb{R}^{m \times 32}$.
For each pair of surface latent, we add a positional encoding of the surface order (whether the first or second surface) in each pair to ensure the intersection is also conditioned on the order of the surfaces.
Then, a set of cross-attention layers $I_{ca}: (\mathbb{R}^{32},\mathbb{R}^{32}) \rightarrow (\mathbb{R}^{16})$ is used to further exchange the features between those two surface latent, where the latent vector of the first surface is treated as the query and the second surface as the key and value.
Finally, we use an MLP $I_{mlp}: \mathbb{R}^{16} \rightarrow \mathbb{R}^{16}$ to map the exchanged features and to the corresponding curve feature. 
We also train a binary classifier $I_{bc}: \mathbb{R}^{16} \rightarrow \{0,1\}$ to identify whether the two surfaces intersect or not.

\paragraph{Decoding and loss function:}
Given the sampled latent $z_s$ and the intersected curve feature $z_c$, we use a decoder network, which consists of CNNs and upsample layers to reconstruct the surface and curve primitives $D: (\mathbb{R}^{m \times 32},\mathbb{R}^{n \times 16}) \rightarrow (\mathbb{R}^{m \times 16 \times 16 \times 3},\mathbb{R}^{n \times 16 \times 3})$:
\begin{align}
    \hat{S_i} = D_{s}(z_s), \hat{C_j} = D_{c}(z_c),
\end{align}
where $\hat{S_i}$ and $\hat{C_j}$ are the reconstructed surface and curve primitives, respectively.

Our loss function consists of three parts: 1) an $L1$ reconstruction loss $\mathcal{L}_{recon}$ of the sample points of surfaces and curves, 2) a binary cross-entropy loss $\mathcal{L}_{inter}$ of the intersection classifier, and 3) a KL regularization loss $\mathcal{L}_{reg}$ to enforce the latent vector to follow a standard Gaussian distribution.

\begin{align}
    \mathcal{L} = w_1\mathcal{L}_{recon} + w_2\mathcal{L}_{inter} + w_3\mathcal{L}_{reg},
\end{align}
\begin{align}
    \mathcal{L}_{recon} = \sum_{i=1}^{m}||S_i - \hat{S_i}||_1 + \sum_{j=1}^{n}||C_j - \hat{C_j}||_1,
\end{align}
\begin{align}
    \mathcal{L}_{inter} = \sum_{i=1}^{m}\sum_{j=1}^{n}T_{SC}(i,j)\log(I_{bc}({S_i}^{(i)},z_s^{(j)})) + \\ 
 (1-T_{SC}(i,j))\log(1-I_{bc}(z_s^{(i)},z_s^{(j)})),
\end{align}
\begin{align}
    \mathcal{L}_{reg} = \sum_{i=1}^{m}KL(\mathcal{N}(0,1)||\mathcal{N}(\mu_i,\sigma_i)),
\end{align}
where $w_1=1,w_2=1e-1,w_3=1e-6$ is the weight of each loss.

\paragraph{Half-Curve Structure:}

During our experiments, we observed that the half-curve structure is critical for robust network training. 
\begin{wrapfigure}{r}{0.25\textwidth}
    \vspace{-0.5cm}
    \hspace{-0.5cm}
    \centering
    \includegraphics[width=0.25\textwidth]{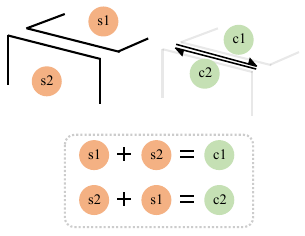} 
    \label{fig:half-curve} 
    \vspace{-0.5cm}
\end{wrapfigure}
While the $16 \times 3$ sample points on a curve inherently have an orientation, directly applying an $L1$ loss between the reconstructed and input curves without considering the half-curve structure causes the network to learn curves with a fixed orientation, which is randomly assigned during data preparation.
An alternative approach is to use the Chamfer distance instead of the $L1$ loss; however, this method is computationally more expensive. Moreover, the curve's orientation is essential for forming curve loops that trim the surface accurately. To address this, we train the network to predict oriented curves (half-curves) based on the order of the surface pair.
The ground truth (GT) half-curve is defined according to its orientation relative to the \textit{first} surface in the surface pair. Specifically, it should form a counter-clockwise loop when serving as the outer boundary of a surface and a clockwise loop when representing an inner hole. If the order of the surface pair is swapped, the predicted half-curve's orientation must also be inverted, as it is now associated with the second surface in the pair.

\subsection{Generation}
\label{sec:method-generation}

\begin{figure}
  \centering
  \includegraphics[width=\linewidth]{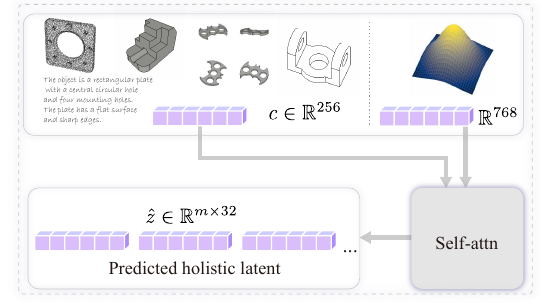}
  \caption{
 Based on the learned latent representation $z_s$, we train a latent diffusion model (LDM) to generate B-Rep models conditioned on various inputs.
 Condition data is processed by fixed or learned feature extractors to produce a $256$-dimensional condition vector $c$.
 Then we use a set of self-attention layers to map the noise to the latent vector $\hat{z}$.
 }
  \label{fig:generation}
\end{figure}

Based on learned latent representation $z_s \in \mathbb{R}^{m \times 32}$, we train a latent diffusion model (LDM) to generate B-Rep models conditioned on various inputs, as shown in Fig.~\ref{fig:generation}.
We first pad the latent vector $z_s$ to a fixed length $z \in \mathbb{R}^{M \times 32}$ by randomly repeating the latent vector $z_s$ until reaching a predefined maximum number of surface primitives $M$.
Then we employ a standard transformer architecture $LDM: (\mathbb{R}^{M \times 32},\mathbb{R}^{256}) \rightarrow \mathbb{R}^{M \times 32}$ to denoise a random noise $\mathcal{N}$ to the latent vector $\hat{z}$, conditioned on a 256-d input $c$.
\begin{align}
    \hat{z} = \mathbb{LDM}(\mathcal{N},c),
\end{align}
where the condition vector $c \in \mathbb{R}^{256}$ is a 256-d feature vector according to the input condition.
The condition vector is set to $0$ when no condition is provided.
We extract other conditions as follows:

\paragraph{Single/multi-view images and sketch:}
For each input image, we use a pre-trained DINOv2~\cite{oquab2024dinov} model to extract a $1024$-dimensional feature vector.
If we have multiple images, we add a positional encoding to encode the view pose information and simply average the feature vector of all images.
Then, we use an MLP layer to map the feature vector to a $256$-dimensional condition vector $c$.
Note that the DINOv2 model is frozen during the training.

\paragraph{Sparse/dense point clouds:}
For each input point cloud, we use a PointNet~\cite{pointnetpp} to extract a $1024$-dimensional feature vector.
Then, we use an MLP layer to map the feature vector to a $256$-dimensional condition vector $c$.
The PointNet model is also trained during the training.

\subsection{Implementation Details}
\label{sec:method-implementation}

For the holistic VAE, we use a batch size of $32$ and train the model on 8 RTX 4090 GPUs with an Adam optimizer and a learning rate of $1e-4$ until the validation loss is converged (around 1M iterations).
We noticed that the number of intersection pairs is much smaller than all possible face pairs, so we use a balanced sampling strategy to ensure the number of positive and negative pairs is equal during the training.
During the inference, we use the intersection classifier to predict the intersection of all pairs.

For the LDM, we use DDPM~\cite{ddpm20} with linear scheduler, beta range of $[0.0001, 0.02]$, a diffusion step of $1000$ and Adam optimizer with a fixed learning rate $1e-4$.
The LDM is trained on 8 RTX 4090 GPUs with a batch size of $128$ for 1M iterations.
The loss is simply set to the $L2$ loss between the predicted latent vector $\hat{z}$ and the ground-truth latent vector $z$.

\paragraph{Model size compared to previous methods:}
Our model is significantly more streamlined compared to prior approaches. BRepGen employs two separate VAEs (88.6M parameters) and four diffusion modules (totaling 202M), each trained and fine-tuned independently. These models are tightly coupled, with later modules depending on the outputs of earlier ones, resulting in a complex interdependency across stages. This cascaded structure not only increases the overall model size but also introduces considerable training and inference complexity. In contrast, our method adopts a unified architecture with a single VAE (181M) and a single diffusion model (205M), enabling end-to-end training within a shared latent space. Despite a larger VAE, our approach avoids the overhead of managing and coordinating multiple dependent components, leading to a more efficient and robust pipeline.

\subsection{Post Processing}
\label{sec:method-post}
Once we obtain the denoised latent vector $\hat{z}$, we feed it into the intersection module $I$ and decoder network $D$ to generate the final B-Rep model $(S,C,T_{SC})$.
Similar to previous methods like SolidGen~\cite{solidgen23} and BRepGen~\cite{brepgen24}, the output is converted into a watertight CAD model using OpenCascade~\cite{opencascade}.
For each surface and curve, we first fit a BSpline primitive with $C_2$ smoothness based on the recovered sample points.
Then, we detect the common endpoints of curves on the same surface and connect them to a wire loop.
Note that each surface can have multiple wire loops, as the largest wire defines the outer boundary of the surface and the rest define the inner holes.
Those oriented wire loops are then used to trim the recovered surface.
Finally, we sew all the surfaces to form a watertight B-Rep model.

\paragraph{Test time augmentation:}
The generative model introduces randomness and diversity in the generated boundary representation (B-Rep) models, even with set conditions. To further enhance the quality of these B-Rep models, we also implement test time augmentation.
For a given point cloud, we run the generative model multiple times, applying different random noise each time to produce several B-Rep models. We then calculate the Chamfer distance between each generated B-Rep model and the input conditions, selecting the model with the smallest distance as the final output.
Since the denoising and post-processing steps can run in parallel, the test time augmentation does not significantly increase the inference time.

\section{Experimental Results}
\label{sec:results}
We present the evaluation of unconditional B-Rep generation in Sec.~\ref{sec:results_unconditional} and conditional B-Rep generation, including point-conditioned in Sec.~\ref{sec:results_points}, text-conditioned in Sec.~\ref{sec:results_text}, image-conditioned (multi-view, single-view, sketch-based) in Sec.~\ref{sec:results_images}. A summary of the limitation is provided in Sec.~\ref{sec:results_limitation}.

\subsection{Unconditional B-Rep Generation}
\label{sec:results_unconditional}

\paragraph{Dataset and baselines.}
We evaluate our method on the commonly used DeepCAD dataset~\cite{deepcad21,ABC} with the official train/val/test split.
Similar to previous methods~\cite{solidgen23,brepgen24}, duplicate models are removed by comparing pairwise light field distance (LFD)~\cite{lfd}.
Additionally, we observed an over-representation of simple shapes (e.g., cubes, cylinders) in both the dataset and the generated samples from prior methods~\cite{deepcad21,brepgen24}.
So we filter out simple models (less than 7 surfaces) and over-complex models (more than 30 surfaces) to better evaluate the quality of the generated samples, which results in 53,225 models in total, including 47,284 in the training set, 3,517 in the validation set, and 2,424 in the test set.
This dataset is used in all our experiments, including both unconditional and conditional generation.
We compare the unconditional generation performance with the state-of-the-art BRepGen~\cite{brepgen24} and DeepCAD~\cite{deepcad21}.

\begin{table*}[t]
    \centering
    \begin{tabular}{l|c|c|c|c|c|c|c|c|c}
        \toprule
                                 & Method                   & Val. ↑            & LFD ↑          & Cov. ↑           & MMD ↓            & JSD ↓            & CC ↑           & MC ↑          \\
        \midrule
        \multirow{3}{*}{DeepCAD} & DeepCAD~\cite{deepcad21} & 50.82\%           & \textbf{1,588} & 69.27\%          & 0.01249          & 0.01159          & 13.53          & 0.62          \\
                                 & BRepGen~\cite{brepgen24} & 47.74\%           & 1,128          & 70.61\%          & 0.01119          & 0.00639          & 11.23          & 1.24          \\
                                 & Ours                     & \textbf{82.68\% } & 1,408          & \textbf{78.87\%} & \textbf{0.01069} & \textbf{0.00536} & \textbf{14.22} & \textbf{2.07} \\
        \midrule
        \midrule
        \multirow{2}{*}{ABC}     & BRepGen~\cite{brepgen24} & 32.68\%           & 1570              & 67.35\%          & 0.01120          & 0.00982          & 13.14          & 1.77          \\
                                 & Ours                     & \textbf{60.46\% } & \textbf{1964}              & \textbf{73.83\%} & \textbf{0.01098} & \textbf{0.00694} & \textbf{19.78} & \textbf{2.93} \\
        \bottomrule
    \end{tabular}
    \caption{
            Qualitative results of unconditional generation on DeepCAD (top) and ABC (bottom) dataset by a) DeepCAD, b) BRepGen, c) ours.
            Numbers are averaged over 10 runs, where each run samples 3000 models from each method and compares it with 1000 models from the ground truth test set.
    }
    \label{tab:quantitative_unconditional}
\end{table*}

\begin{figure*}[t]
    \centering
    \includegraphics[width=0.84\linewidth]{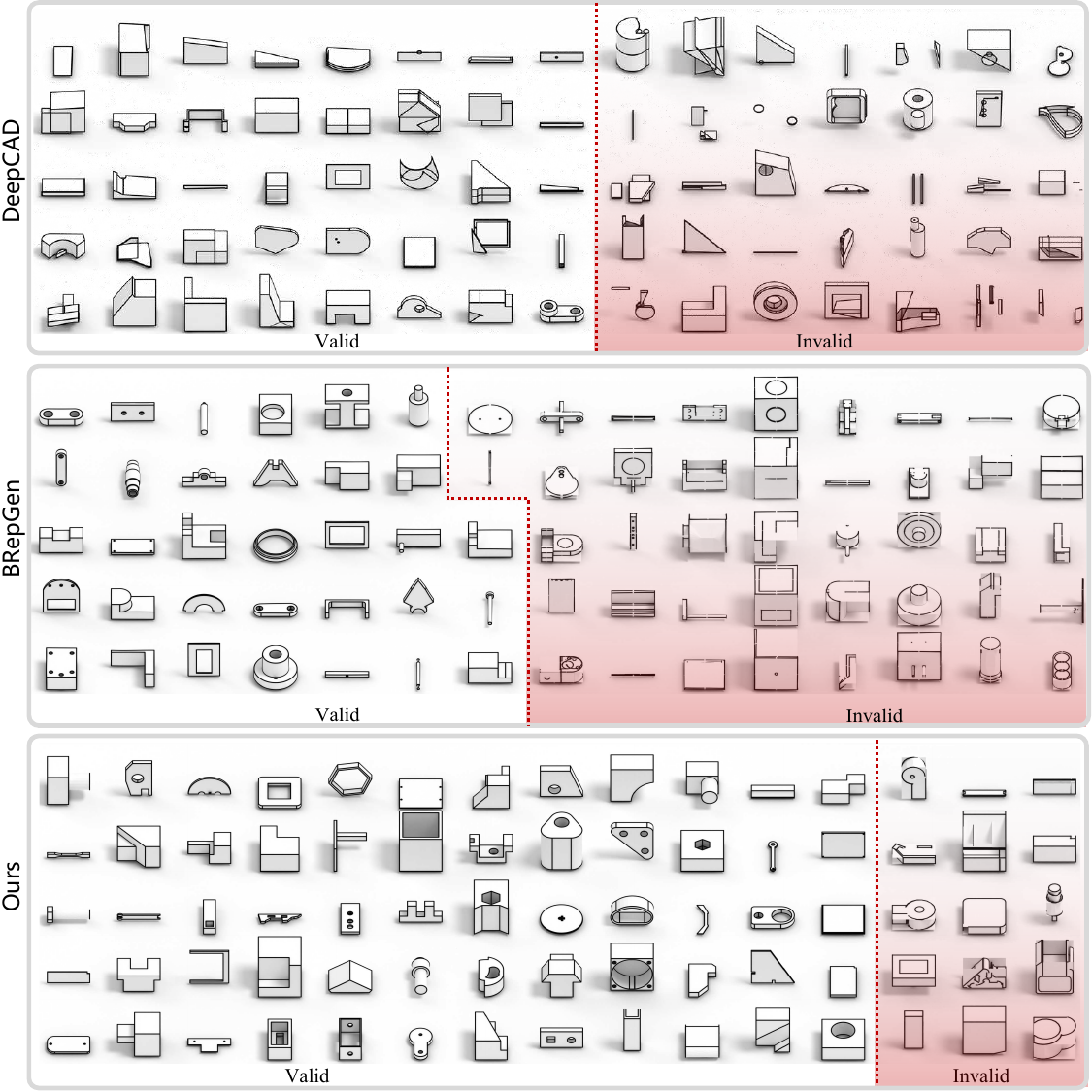}
    \caption{
        Qualitative results of unconditional generation on DeepCAD dataset by a) DeepCAD, b) BRepGen, c) ours. Models are \textbf{randomly} selected from the generated samples after filtering out simple models (less than 7 surfaces). Valid and invalid models are split by the dashed line.
    }
    \label{fig:qualitative_unconditional}
\end{figure*}

\begin{figure*}
    \centering
    \includegraphics[width=1\linewidth]{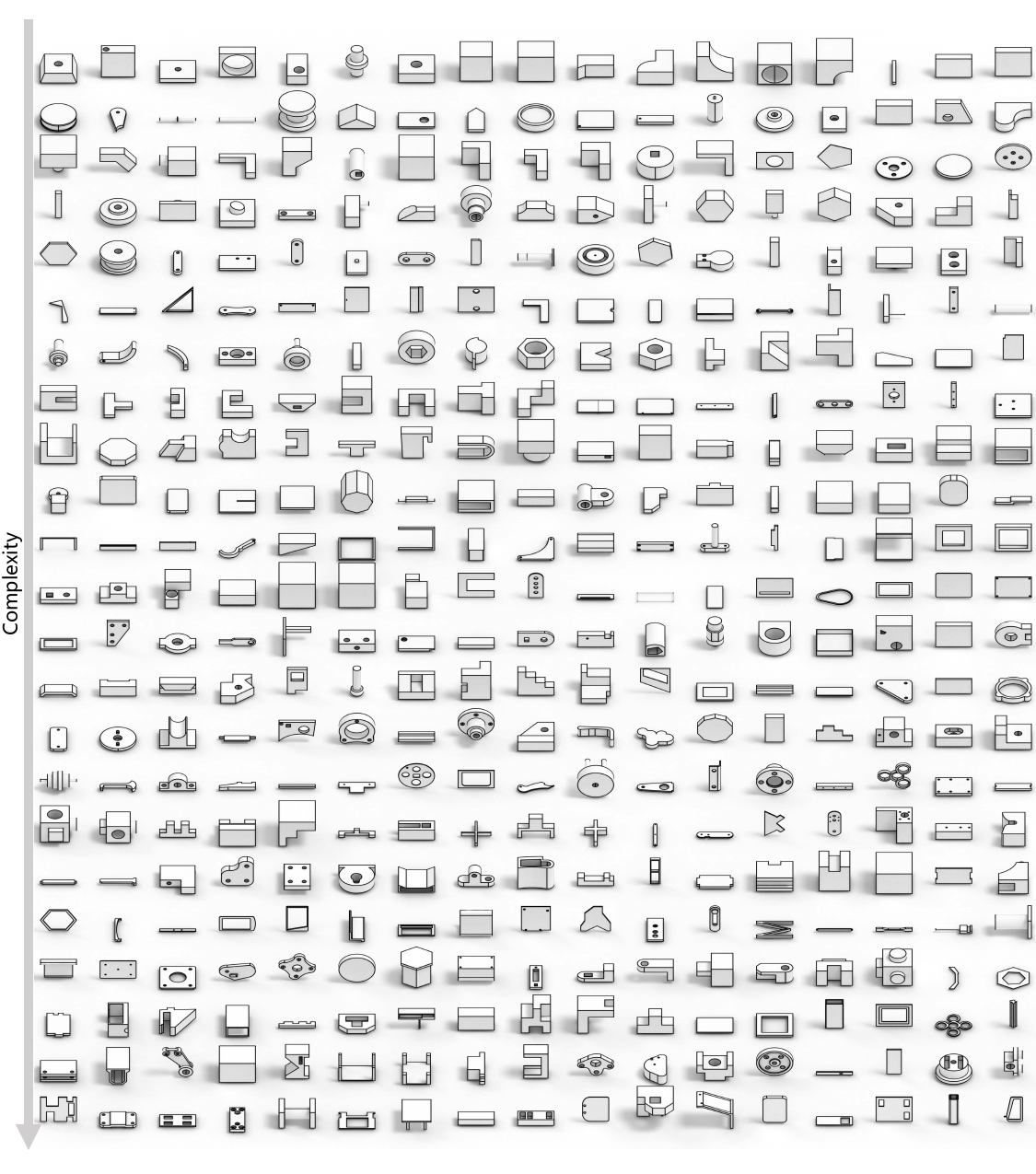}
    \caption{
        Qualitative results of our unconditional generation result as complexity increases.
    }
    \label{fig:qualitative_unconditional2}
\end{figure*}

\begin{figure*}
    \centering
    \includegraphics[width=1\linewidth]{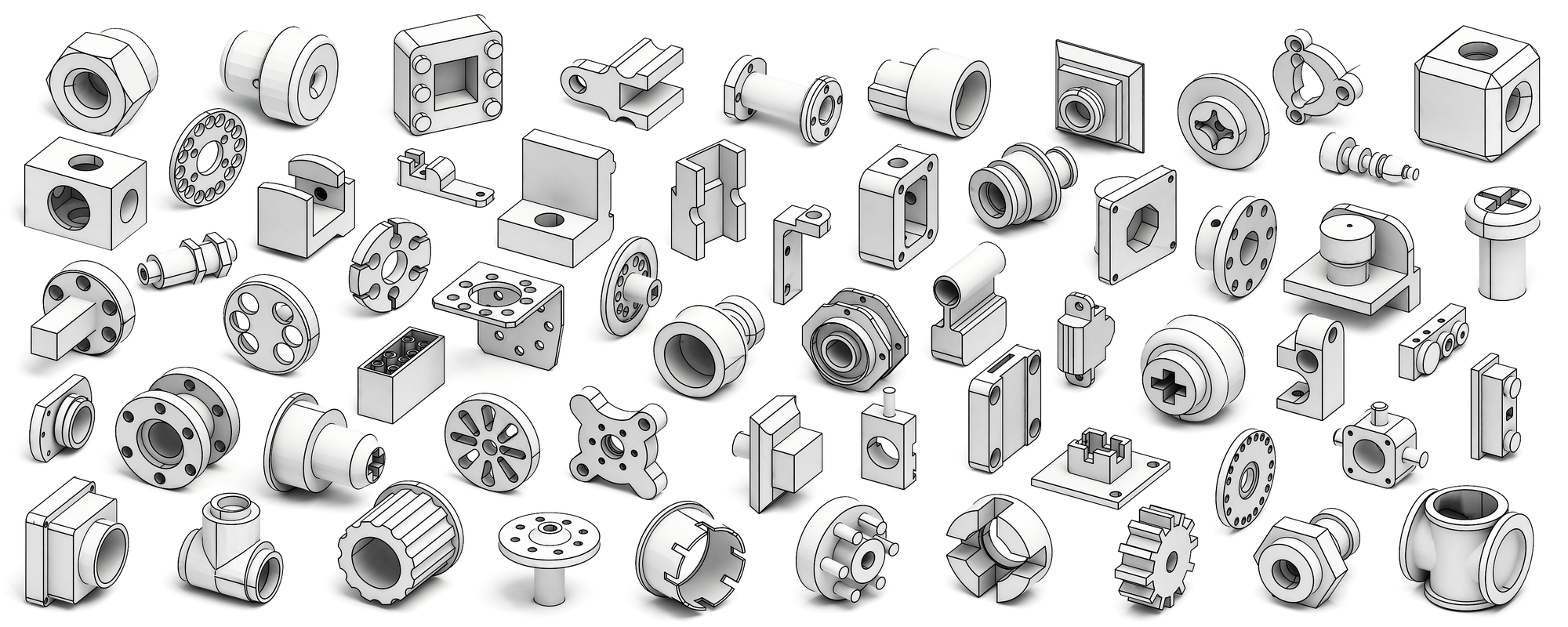}
    \caption{
        Qualitative results of our unconditional generation on ABC dataset.
    }
    \label{fig:qualitative_unconditional_abc}
\end{figure*}

\begin{figure}
    \centering
    \includegraphics[width=1\linewidth]{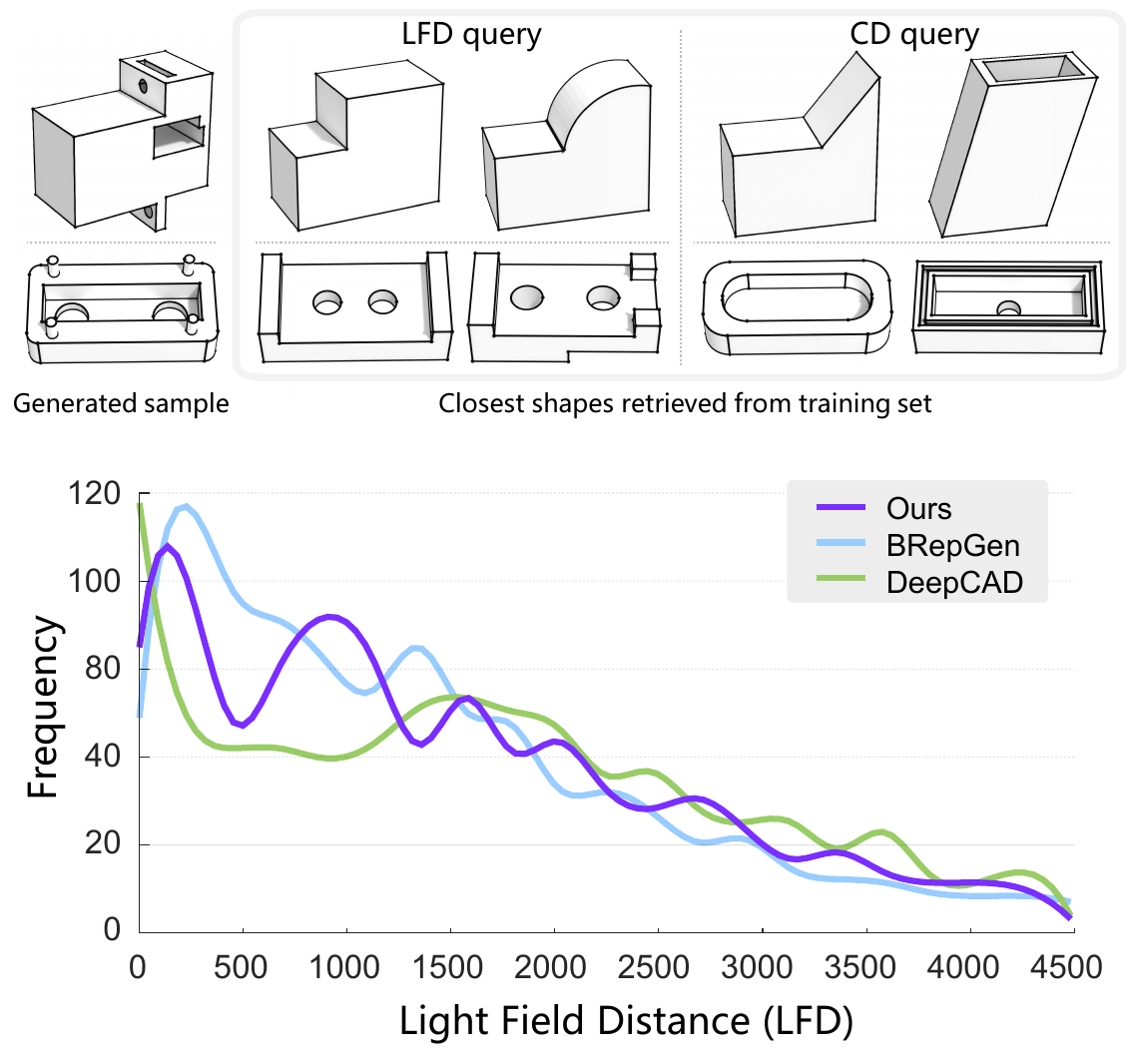}
    \caption{
        Novelty analysis of unconditional generation in terms of the light field distance from the generated samples to the DeepCAD training set by a) DeepCAD, b) BRepGen, and c) ours.
        Lower light field distance indicates the generated samples are more similar to the training set (less novel).
    }
    \label{fig:novel_unconditional}
\end{figure}

\paragraph{Metrics.}
We follow a similar evaluation protocol as previous works~\cite{solidgen23,brepgen24} to evaluate the quality of the generated samples by the following metrics:
\begin{itemize}
    \item \textbf{Validity (Val.) ↑} measures the ratio of valid models in the generated samples.
    \item \textbf{Averaged Light Field Distance (LFD) ↑} measures the novelty of the generated samples by comparing the light field distance between the generated samples and the whole training set. This metric is critical to prevent the model from overfitting to the training set~\cite{lfd}.
    \item \textbf{Coverage(Cov.) ↑} measures the ratio of the test set that are covered by the generated samples, more specifically, is the percentage of test set with at least one match after assigning every generated data to its closest neighbor in the test set based on Chamfer Distance (CD).
    \item \textbf{Jensen-Shannon Divergence (JSD) ↓} measures the distribution difference between the generated samples and the test set after converting point clouds into $28^3$ discrete voxels.
    \item \textbf{Maximum Mean Discrepancy (MMD) ↓} measures the averaged CD between the each generated sample and its nearest neighbor in the test set.
    \item \textbf{Cyclomatic Complexity (CC) ↑} measures the complexity of the generated samples by counting the number of loops in the wireframe graph representation of the B-Rep model~\cite{Cyclomatic23}.
    \item \textbf{Mean curvature (MC) ↑} measures the averaged mean curvature of the generated samples.
\end{itemize}
\textit{Val.} assesses the model's ability to generate valid B-Reps.
As for distribution-based metrics, \textit{LFD} evaluates novelty against the training set, while \textit{Cov.}, \textit{MMD} and \textit{JSD} measure the plausibility relative to the test set.

We collect 3,000 models per method for comparison against the DeepCAD test set. For BRepGen and our method, we sample 8,000 models and filter out simple models (fewer than 7 surfaces). Due to its low valid ratio and limited model complexity, 20,000 samples are required for DeepCAD to achieve a comparable number of shapes. We evaluate validity using the BRep validity checker from OpenCascade~\cite{opencascade} with a precision of 0.1. For distribution metrics (\textit{Cov.}, \textit{MMD}), we randomly sample 1,000 test set models and report averaged results over 10 runs.

\paragraph{Discussion.}
Table~\ref{tab:quantitative_unconditional} shows the quantitative results for unconditional generation across DeepCAD, BRepGen, and our method. BRepGen achieves an initial valid ratio of 62.9\%, which decreases to 47.74\% after filtering out simple models. In contrast, our method achieves a significantly higher valid ratio of 82.68\%, showcasing the effectiveness of fusing geometry and topology within a holistic latent space.

Although DeepCAD shows a higher LFD distance, the generated shapes are less complex and plausible than those of other methods. Its sketch-and-extrude process biases the results toward planar shapes, as indicated by its low mean curvature in Table~\ref{tab:quantitative_unconditional}, limiting the diversity of generated samples. Furthermore, most of DeepCAD's generated samples involve only 1 or 2 extrude operations, indicating that the model does not effectively capture the transition in the sketch-and-extrude process.
Qualitative results in Fig.~\ref{fig:qualitative_unconditional} further demonstrate our method's superiority. Among 75 randomly selected generated models, our method produces more diverse and plausible shapes compared to the baselines, with valid and invalid models separated by a dashed line. Fig.~\ref{fig:novel_unconditional} illustrates the distribution of LFD distances to the training set, where our method achieves balanced results across all samples. A complete gallery of generated samples is provided in Fig.~\ref{fig:qualitative_unconditional2} for DeepCAD dataset and Fig.~\ref{fig:qualitative_unconditional_abc} for ABC dataset, highlighting the diversity of our models.

\paragraph{Failures.}
Fig.~\ref{fig:qualitative_unconditional} includes 15 failure cases from our method. These failures primarily result from inconsistently generated surface primitives, an issue observed in previous methods as well. Although our approach encodes geometry and topology in a holistic latent space, the latent space is constructed from several surface primitives. Consequently, inconsistencies in surface primitives still lead to inaccurate trimming and non-watertight B-Rep models.

\subsection{Point Conditioned B-Rep Generation}
\label{sec:results_points}

    \begin{table}
        \centering
        \resizebox{\columnwidth}{!}{
            \begin{tabular}{l|c|c|c|c|c|c|c|c|c}
                \toprule
                Method             & \multicolumn{3}{c|}{Chamfer ↓} & \multicolumn{3}{c|}{Geometry ↑} & \multicolumn{2}{c|}{Topology ↑} & \multirow{2}{*}{Val.}                                                                           \\
                \cmidrule(r){2-4} \cmidrule(r){5-7} \cmidrule(r){8-9}
                Name               & Vertex                         & Edge                            & Face                            & Vertex                & Edge          & Face          & FE            & EV            &         \\
                \midrule
                HPNet+P2C          & 0.1676                         & 0.0146                          & 0.0031                          & 0.72                  & 0.78          & 0.83          & 0.75          & 0.64          & /       \\
                SEDNet+P2C         & 0.1534                         & 0.0082                          & 0.0024                          & 0.74                  & 0.78          & 0.81          & 0.73          & 0.63          & /       \\
                Ours               & 0.0474                         & 0.0054                          & 0.0029                          & 0.86                  & 0.89          & 0.94          & 0.86          & 0.84          & 83.87\% \\
                Ours               & \textbf{0.0424}                & \textbf{0.0028}                 & \textbf{0.0008}                 & \textbf{0.89}         & \textbf{0.92} & \textbf{0.96} & \textbf{0.91} & \textbf{0.88} & \textbf{98.23\%} \\
                \midrule
                NVD-Net            & 0.0372                         & 0.0091                          & 0.0008                          & 0.90                  & 0.90          & 0.89          & 0.84          & 0.84          & /       \\
                Ours               & 0.0133                         & 0.0039                          & 0.0029                          & 0.93                  & 0.93          & 0.95          & 0.90          & 0.89          & 86.11\%       \\
                $\text{Ours}_{32}$ & \textbf{0.0091}                & \textbf{0.0015}                 & \textbf{0.0007}                 & \textbf{0.95}         & \textbf{0.95} & \textbf{0.96} & \textbf{0.93} & \textbf{0.93} & \textbf{98.62\%}       \\
                \bottomrule
            \end{tabular}
        }
        \caption{
                Quantitative results of point-conditioned generation on DeepCAD by a) HPNet+Point2CAD, b) SEDNet+Point2CAD, c) NVD-Net, d) ours.
        }
        \label{tab:quantitative_points}
    \end{table}

\begin{table}
    \centering
    \resizebox{\columnwidth}{!}{
        \begin{tabular}{l|c|c|c|c|c|c|c|c|c}
            \toprule
            Method                & \multicolumn{3}{c|}{Chamfer ↓} & \multicolumn{3}{c|}{Geometry ↑} & \multicolumn{3}{c}{Topology ↑}                                                                                 \\
            \cmidrule(r){2-4} \cmidrule(r){5-7} \cmidrule(r){8-9}
            Name                  & Vertex                         & Edge                            & Face                           & Vertex        & Edge          & Face          & FE            & EV            \\
            \midrule
            Base                  & \textbf{0.0494}                & \textbf{0.0060}                 & \textbf{0.0035}                & \textbf{0.86} & \textbf{0.90} & \textbf{0.94} & \textbf{0.87} & \textbf{0.85} \\
            +\textit{w/ cropping} & 0.0698                         & 0.0089                          & 0.0037                         & 0.80          & 0.83          & 0.88          & 0.82          & 0.79          \\
            +\textit{w/ noise}    & 0.0652                         & 0.0085                          & 0.0036                         & 0.81          & 0.84          & 0.89          & 0.83          & 0.81          \\
            +\textit{w/o normal}  & 0.0836                         & 0.0121                          & 0.0048                         & 0.75          & 0.78          & 0.83          & 0.78          & 0.75          \\
            +\textit{w/ sparsify} & 0.0993                         & 0.0156                          & 0.0066                         & 0.72          & 0.74          & 0.80          & 0.74          & 0.72          \\
            \bottomrule
        \end{tabular}
    }
    \caption{
        Quantitative results of point-conditioned generation with imperfect input data.
        Our method can still achieve comparable performance with different types of noise and data imperfections.
    }
    \label{tab:quantitative_points_noise}
\end{table}

\begin{table*}
    \centering
    \resizebox{\linewidth}{!}{
        \begin{tabular}{l|c|c|c|c|c|c|c|c|c|c|c|c|c|c|c|c|c|c|c}
            \toprule
            Method             & \multicolumn{3}{c|}{Number} & \multicolumn{3}{c|}{Accuracy Error ↓} & \multicolumn{3}{c|}{Completeness Error↓} & \multicolumn{5}{c|}{Precision ↑} & \multicolumn{5}{c}{Recall ↑}                                                                                                                                                                                                                                             \\
            \cmidrule(r){2-4} \cmidrule(r){5-7} \cmidrule(r){8-10} \cmidrule(r){11-15} \cmidrule(r){16-20}
            Name               & Vertex                      & Edge                                  & Face                                     & Vertex                           & Edge                         & Face             & Vertex           & Edge             & Face             & Vertex        & Edge          & Face          & FE            & EV            & Vertex        & Edge          & Face          & FE            & EV            \\
            \midrule
            HPNet+P2C          & 19/21                       & 25/31                                 & 11/13                                    & 0.0322                           & 0.0017                       & 0.0029           & 0.1354           & 0.0129           & \textbf{0.0002}  & 0.84          & 0.88          & 0.89          & 0.83          & 0.72          & 0.69          & 0.74          & 0.79          & 0.71          & 0.61          \\
            SEDNet+P2C         & 16/21                       & 22/31                                 & 10/13                                    & \textbf{0.0195}                  & \textbf{0.0012}              & 0.0021           & 0.1340           & 0.0070           & 0.0003           & 0.88          & 0.92          & 0.92          & 0.85          & 0.73          & 0.68          & 0.70          & 0.75          & 0.67          & 0.58          \\
            Ours               & 21/21                       & 32/31                                 & 12/13                                    & 0.0220                           & 0.0024                       & 0.0014           & 0.0255           & 0.0030           & 0.0016           & 0.88          & 0.89          & 0.96          & 0.88          & 0.85          & 0.87          & 0.91          & 0.93          & 0.85          & 0.85          \\
            $\text{Ours}_{32}$ & 20/21                       & 30/31                                 & 12/13                                    & 0.0197                           & 0.0013                       & \textbf{0.0005}  & \textbf{0.0227}  & \textbf{0.0015}  & 0.0003           & \textbf{0.91} & \textbf{0.93} & \textbf{0.98} & \textbf{0.93} & \textbf{0.90} & \textbf{0.87} & \textbf{0.91} & \textbf{0.95} & \textbf{0.90} & \textbf{0.88} \\
            \midrule
            NVD-Net            & 20/20                       & 28/31                                 & 11/13                                    & 0.00744                          & 0.00454                      & 0.00019          & 0.02976          & 0.00453          & 0.00061          & 0.94          & \textbf{0.97} & 0.96          & 0.92          & 0.90          & 0.90          & 0.87          & 0.86          & 0.81          & 0.83          \\
            Ours               & 22/20                       & 32/31                                 & 12/13                                    & 0.00714                          & 0.00198                      & 0.00157          & 0.00620          & 0.00195          & 0.00131          & 0.93          & 0.92          & 0.96          & 0.91          & 0.89          & 0.95          & 0.95          & 0.94          & 0.89          & 0.9           \\
            $\text{Ours}_{32}$ & 20/20                       & 30/31                                 & 12/13                                    & \textbf{0.00430}                 & \textbf{0.00070}             & \textbf{0.00049} & \textbf{0.00477} & \textbf{0.00077} & \textbf{0.00023} & \textbf{0.97} & 0.95          & \textbf{0.98} & \textbf{0.95} & \textbf{0.94} & \textbf{0.94} & \textbf{0.95} & \textbf{0.95} & \textbf{0.93} & \textbf{0.92} \\
            \bottomrule
        \end{tabular}
    }
    \caption{
            Detailed quantitative results of point-conditioned generation on DeepCAD by a) HPNet+Point2CAD, b) SEDNet+Point2CAD, c) NVD-Net, d) ours.
            The table presents the number of vertices, curves, and faces in the generated samples compared to the ground truth, along with metrics for reconstruction accuracy and completeness of individual primitive geometries.
            We also report the precision and recall of both geometry and topology for generated samples.
    }
    \label{tab:quantitative_detailed}
\end{table*}

\begin{figure*}
    \centering
    \includegraphics[width=1\linewidth]{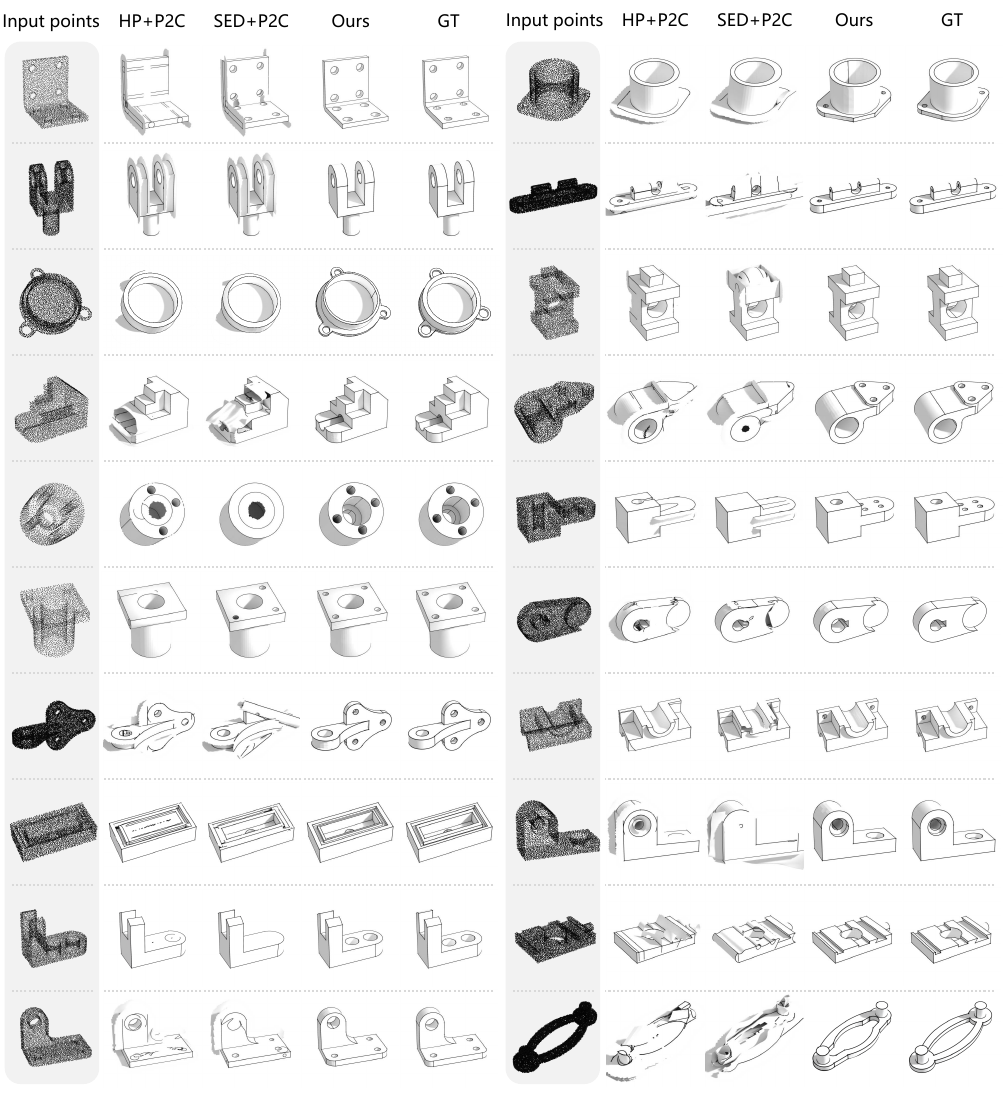}
    \caption{
        Qualitative results of point-conditioned generation on DeepCAD test set by a) HPNet+Point2CAD, b) SEDNet+Point2CAD, c) ours.
        Compared to the other baseline methods, our holistic latent representation facilitates a more consistent relationship between topology and geometry, leading to more plausible generation results.
    }
    \label{fig:qualitative_points}
\end{figure*}

\begin{figure*}
    \centering
    \includegraphics[width=1\linewidth]{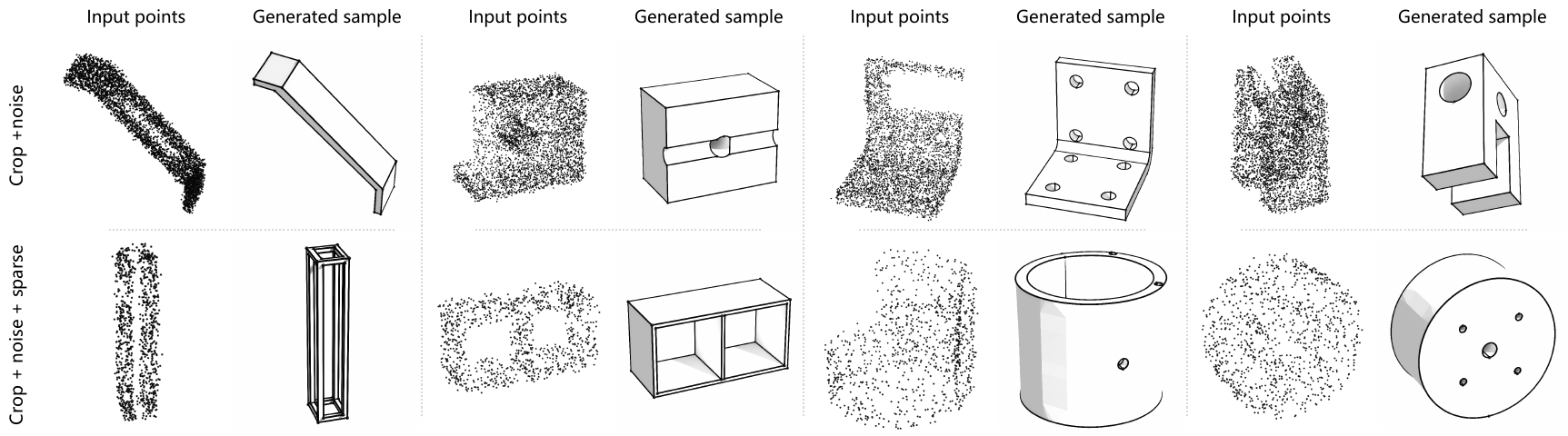}
    \caption{
        Qualitative results of point-conditioned generation on imperfect point clouds, including a) random cropping, b) adding 2\% Gaussian noise, c) masking normal, and d) downsampling to 1,024 points.
    }
    \label{fig:qualitative_points_noise}
\end{figure*}

\begin{figure*}
    \centering
    \includegraphics[width=1\linewidth]{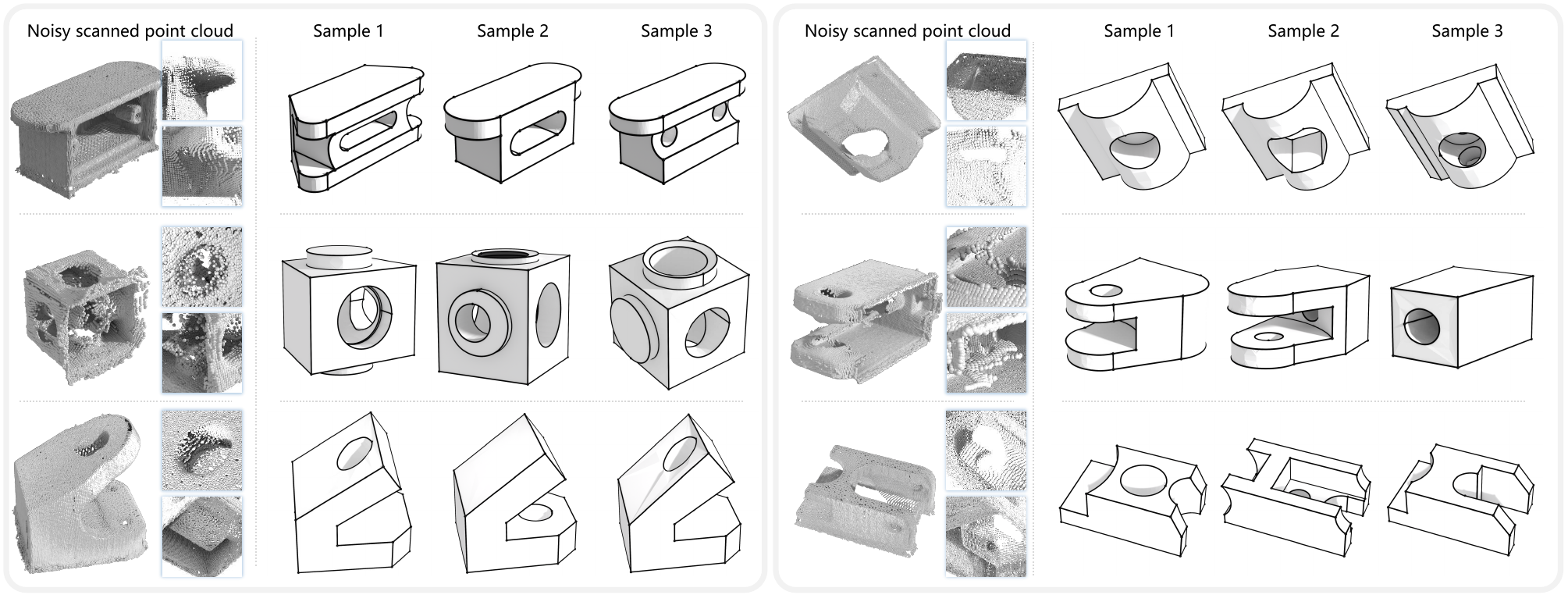}
    \caption{
        Qualitative results of point-conditioned generation on scanned point clouds from the structured light scanner.
        The scanned point cloud usually contains 1) noise, 2) missing parts, and 3) misalignment due to the limitations of scanning devices and the environment.
        However, our method can still generate multiple plausible candidates from the noisy input despite the heavy noise.
    }
    \label{fig:qualitative_points_real_noise}
\end{figure*}

\begin{figure}
    \centering
    \includegraphics[width=1\linewidth]{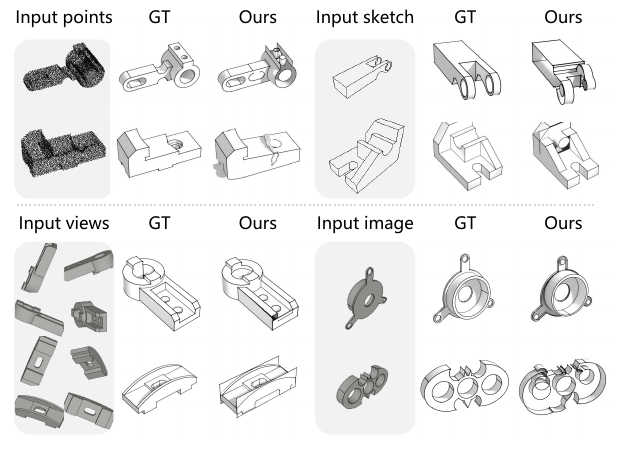}
    \caption{
        Failure cases of conditioned generation.
    }
    \label{fig:qualitative_failure}
\end{figure}

\paragraph{Dataset and Baselines.}
We compare our point-conditioned generation results with three state-of-the-art methods, NVD-Net~\cite{nvd24}, HPNet~\cite{hpnet21}+Point2CAD~\cite{point2cad} and SEDNet~\cite{sed23}+Point2CAD, on the same DeepCAD test set as above. The result for NVD-Net~\cite{nvd24} is only computed on a subset of the test set (800 shapes) since it only supports elementary primitives. For other baseline methods, we use the official implementations and pre-trained models provided by the authors to generate point segmentations on 10,000 points with normal vectors. We then use the Point2CAD pipeline to reconstruct B-Rep models from the segmented point clouds.
For our method, we use the same point cloud as input and train the diffusion model until convergence on the DeepCAD validation set.
We include two versions of our method: a base version (denoted as \textit{Ours}) which only samples once per input point cloud, and a version with 32 samples per input point cloud (denoted as $\text{Ours}_{32}$) to evaluate the impact of sampling frequency on the generation quality.
For metrics computing, for each model, we randomly sampled points on the surface (10,000 points per square unit) and curves (100 points per unit) based on their area and length. We also extract surface-to-curve (FE) and curve-to-vertex (EV) topological connectivity.

\paragraph{Metrics.}
Following a similar evaluation protocol~\cite{complexgen22,nvd24} as previous works, we first match the predicted primitives with the GT data using Hungarian matching algorithm and compute the following metrics:
\begin{itemize}
    \item \textbf{Chamfer ↓} measures Chamfer Distance (CD) between the generated samples and the ground truth in matched primitive level.
    \item \textbf{Geometry ↑} measures the F-score of matched primitives.
    \item \textbf{Topology ↑} measures the F-score of matched topological connections.
\end{itemize}
A detailed breakdown of the quantitative results is provided in Table~\ref{tab:quantitative_detailed}, including:
\begin{itemize}
    \item \textbf{Number} measures the number of primitives correctly recovered.
    \item \textbf{Accuracy ↓} measures the one-way distance from the matched primitives to the ground truth.
    \item \textbf{Completeness ↓} measures the one-way distance from the ground truth to the matched primitives.
    \item \textbf{Precision ↑} measures the precision of recovered geometry and topological connections (0.1 thresholds).
    \item \textbf{Recall ↑} measures recall of recovered geometry and topological connections (0.1 threshold).
\end{itemize}

\paragraph{Discussion.}
As shown in Table~\ref{tab:quantitative_points}, our method, with a single sample per input point cloud, achieves a comparable Chamfer distance and superior F-scores for geometry and topology compared to the baselines. Additionally, the $\text{Ours}_{32}$ variant outperforms the baselines across all metrics. Notably, our method is also more efficient, with a significantly lower runtime.

The standard version of our method takes 0.56s per model to generate a B-Rep, including 0.27s for denoising, 0.27s for post-processing, and 0.02s for selection. The $\text{Ours}_{32}$ variant requires 12.46s per model, with 8.66s for denoising, 3.26s for post-processing, and 0.54s for selection. In comparison, HPNet and SEDNet both take over 30s per model due to the heavy computational requirements of clustering and fitting processes.

A detailed breakdown of the quantitative results is provided in Table~\ref{tab:quantitative_detailed}. Our method generates a number of primitives closest to the ground truth and outperforms the baselines in most detailed metrics. Interestingly, fitting-based methods (HPNet+Point2CAD, SEDNet+Point2CAD) tend to produce more complete surfaces, as evidenced by the lower completeness error in Table~\ref{tab:quantitative_detailed} and visualized in Fig.~\ref{fig:qualitative_points}. However, the lack of topological information prevents these surfaces from being correctly trimmed, leading to lower reconstruction accuracy and topological scores.
While NVD-Net achieves higher reconstruction accuracy compared to other baselines, our method still outperforms it across most metrics—particularly in topology—due to its explicit modeling of topological structures, which NVD-Net lacks during training.

The high topological errors observed in HPNet+Point2CAD and SEDNet+Point2CAD often result in invalid B-Rep models that cannot be imported into CAD software for further editing. In contrast, our method leverages a holistic latent representation that captures both geometry and topology, resulting in more consistent relationships between these aspects and producing more plausible and valid generation outcomes.

\paragraph{Robustness.}
We conducted robustness tests on the point-conditioned generation task to evaluate the model's performance under noisy and incomplete point clouds. Starting with clean point clouds containing accurate normal vectors, we incrementally introduce: 1) random cropping using an axis-aligned bounding box of $[0-1]$ length, 2) adding 2\% Gaussian noise, 3) masking normal vectors with a 0.5 probability, and 4) downsampling the point cloud to 1,024 points. During training, these operations are randomly applied to improve robustness.
As shown in Table~\ref{tab:quantitative_points_noise} and Fig.~\ref{fig:qualitative_points_noise}, our model maintains comparable performance under imperfect input conditions. Additionally, we evaluate the model on real-world scanned point clouds~\cite{DEF22} from a structured light scanner, shown in Fig.~\ref{fig:qualitative_points_real_noise}. These point clouds often include noise, missing regions, and misalignments due to hardware and environmental limitations. Despite these challenges, our method generates multiple plausible candidates from the noisy input, demonstrating its robustness.

\begin{figure}
    \centering
    \includegraphics[width=0.97\linewidth]{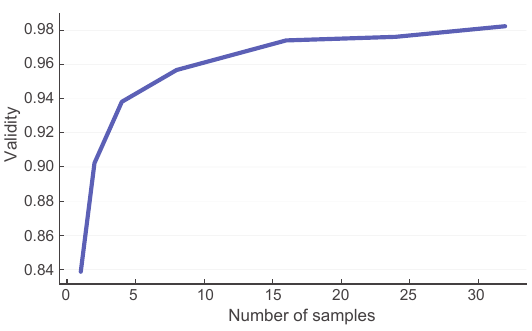}
    \caption{
        Relation between the number of runs and the valid ratio of the point conditioned generation.
    }
    \label{fig:tta}
\end{figure}

\paragraph{Num. of run vs Val. ratio.}
Unlike unconditional generation, the valid ratio of conditional generation can be significantly boosted by various test time augmentation methods.
As shown in Table~\ref{tab:quantitative_points}, the valid ratio of our method can be increased from 83.87\% to 98.23\% by simply increasing the number of runs from 1 to 32.
We also show a plot in Fig.~\ref{fig:tta} to illustrate the relationship between the number of runs and the valid ratio.
Despite the nondeterministic nature, the generative model can still achieve a high valid ratio and lower reconstruction error compared with the deterministic baselines with a simple multi-run strategy.
More advanced test time augmentation methods, like encoding difference between the previous generated shape and the input point cloud to facilitate the future generation, could further enhance the results while lowering computational costs.

\paragraph{Failures.}
Fig.~\ref{fig:qualitative_failure} (upper left corner) shows two failure cases from our method. These failures are primarily caused by inconsistencies in the generated surface primitives, which can lead to incorrect trimming and topological connections.

\begin{figure*}
    \centering
    \includegraphics[width=0.97\linewidth]{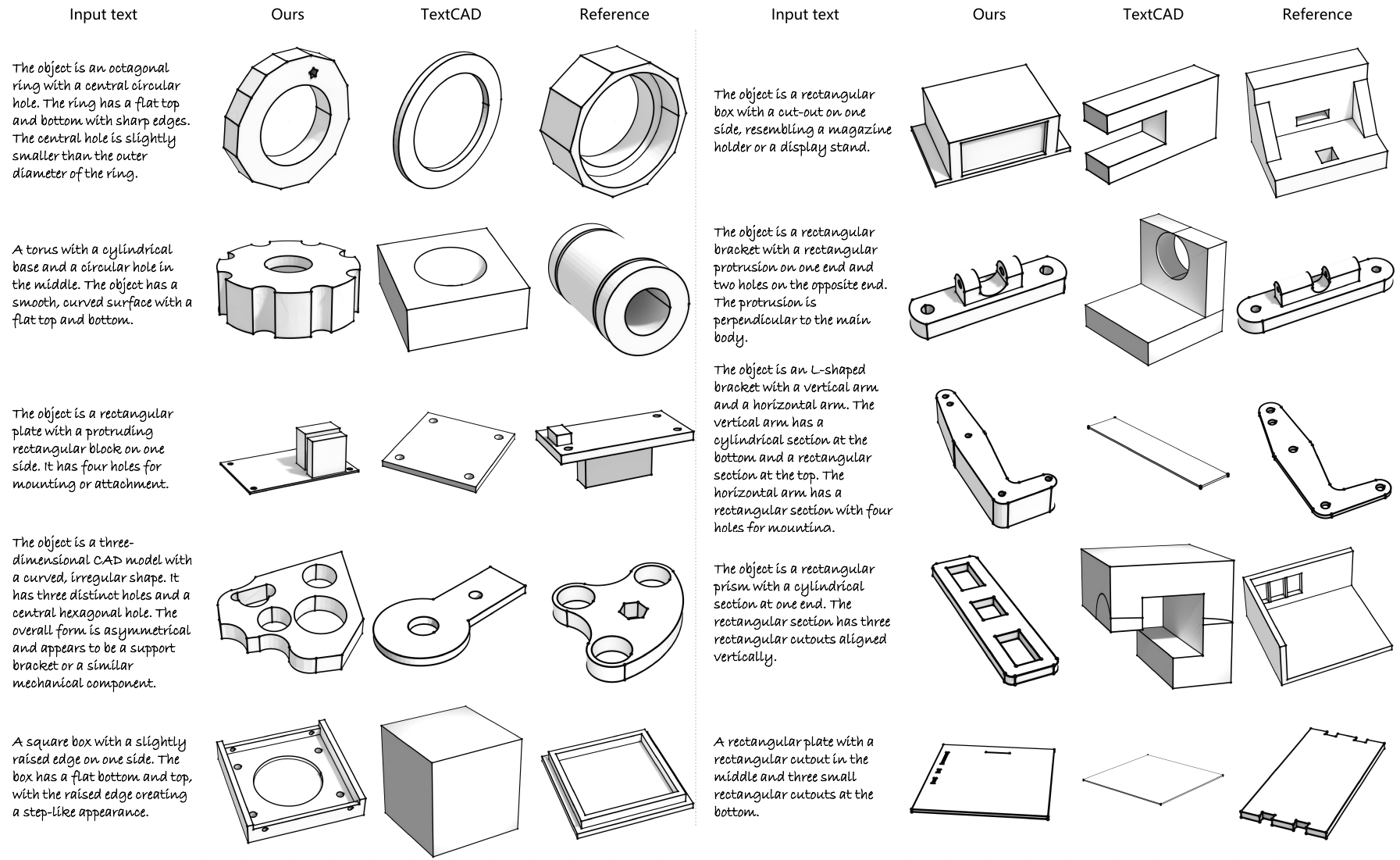}
    \caption{
        Qualitative results of text-based conditioned generation.
        Annotations come from the test set of Text2CAD~\cite{khan2024textcad}.
    }
    \label{fig:qualitative_txt}
\end{figure*}

\begin{figure*}
    \centering
    \includegraphics[width=0.97\linewidth]{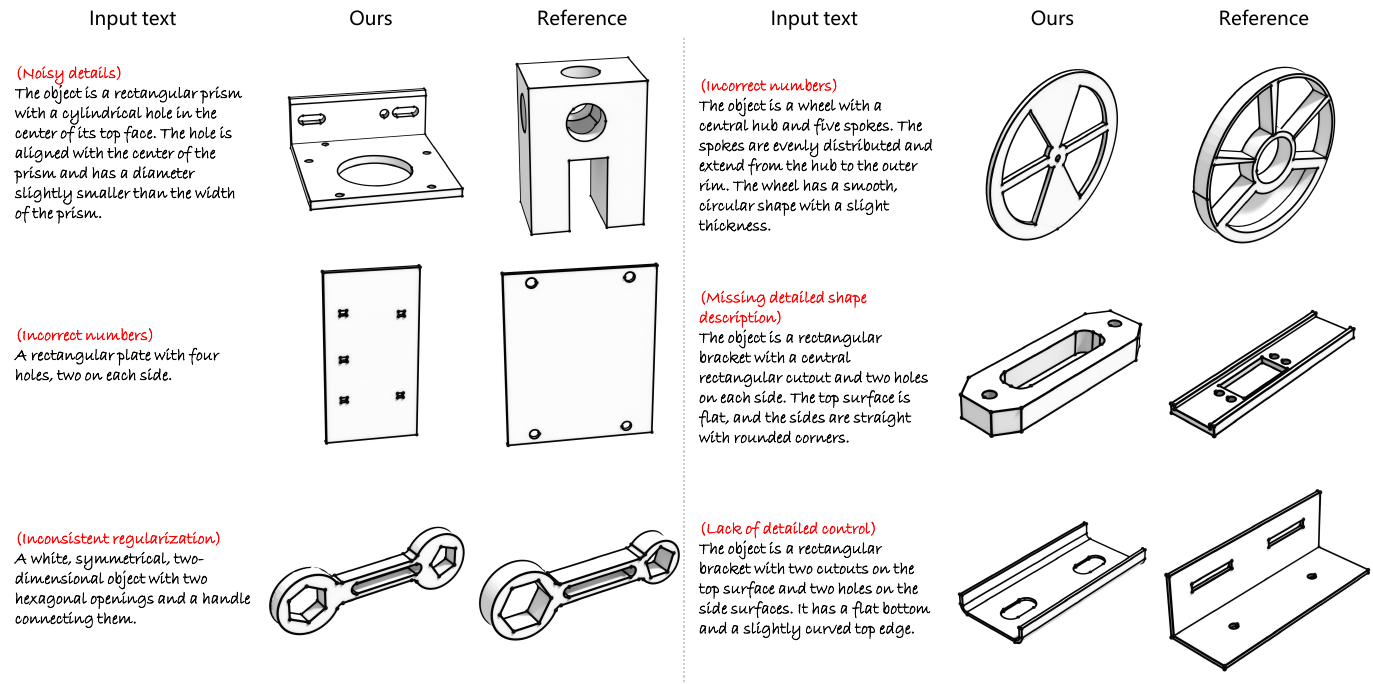}
    \caption{
        Failure cases of txt-based conditioned generation. Primitive numbers, detailed shape description and inconsistent regularization, are the main reasons for the failure.
    }
    \label{fig:qualitative_txt_failed}
\end{figure*}

\begin{table*}
    \centering
    \resizebox{\linewidth}{!}{
        \begin{tabular}{l|c|c|c|c|c|c|c|c|c|c|c|c|c|c|c|c|c|c|c|c}
            \toprule
            Method                                         & \multicolumn{3}{c|}{Number} & \multicolumn{3}{c|}{Accuracy Error ↓} & \multicolumn{3}{c|}{Completeness Error↓} & \multicolumn{5}{c|}{Precision ↑} & \multicolumn{5}{c|}{Recall ↑} & \multirow{2}{*}{Val.}                                                                                                                                                                                                                                          \\
            \cmidrule(r){2-4} \cmidrule(r){5-7} \cmidrule(r){8-10} \cmidrule(r){11-15} \cmidrule(r){16-20}
            Name                                           & Vertex                      & Edge                                  & Face                                     & Vertex                           & Edge                          & Face                  & Vertex          & Edge            & Face            & Vertex        & Edge          & Face          & FE            & EV            & Vertex        & Edge          & Face          & FE            & EV            &                  \\
            \cmidrule(r){1-1} \cmidrule(r){2-4} \cmidrule(r){5-7} \cmidrule(r){8-10} \cmidrule(r){11-15} \cmidrule(r){16-20} \cmidrule(r){21-21}
            Full                                           & 20/21                       & 31/31                                 & 12/13                                    & \textbf{0.0014}                  & 0.0006                        & 0.0006                & 0.0038          & 0.0023          & 0.0020          & \textbf{0.99} & \textbf{0.97} & \textbf{1.00} & \textbf{0.98} & \textbf{0.97} & \textbf{0.98} & \textbf{0.98} & \textbf{0.99} & \textbf{0.97} & \textbf{0.97} & \textbf{94.18\%} \\
            w/o self attn                                  & 21/21                       & 31/31                                 & 12/13                                    & 0.0019                           & \textbf{0.0004}               & \textbf{0.0005}       & \textbf{0.0032} & \textbf{0.0018} & \textbf{0.0016} & 0.98          & 0.97          & 1.00          & 0.97          & 0.96          & 0.97          & 0.98          & 0.98          & 0.95          & 0.96          & 91.63\%          \\
            w/o self attn + spatial resolution             & 21/21                       & 31/31                                 & 12/13                                    & 0.0025                           & 0.0010                        & 0.0009                & 0.0047          & 0.0037          & 0.0034          & 0.97          & 0.96          & 0.99          & 0.96          & 0.95          & 0.97          & 0.97          & 0.97          & 0.94          & 0.94          & 88.37\%          \\
            w/o self attn + spatial resolution + half edge & 22/21                       & 32/31                                 & 12/13                                    & 0.0045                           & 0.0022                        & 0.0019                & 0.0070          & 0.0049          & 0.0040          & 0.94          & 0.94          & 0.99          & 0.93          & 0.9           & 0.96          & 0.97          & 0.96          & 0.9           & 0.91          & 79.62\%          \\
            \bottomrule
        \end{tabular}
    }
    \caption{
            Quantitative results of ablation study on holistic VAE.
    }
    \label{tab:quantitative_ablation}
\end{table*}

\begin{figure}
    \centering
    \includegraphics[width=1\linewidth]{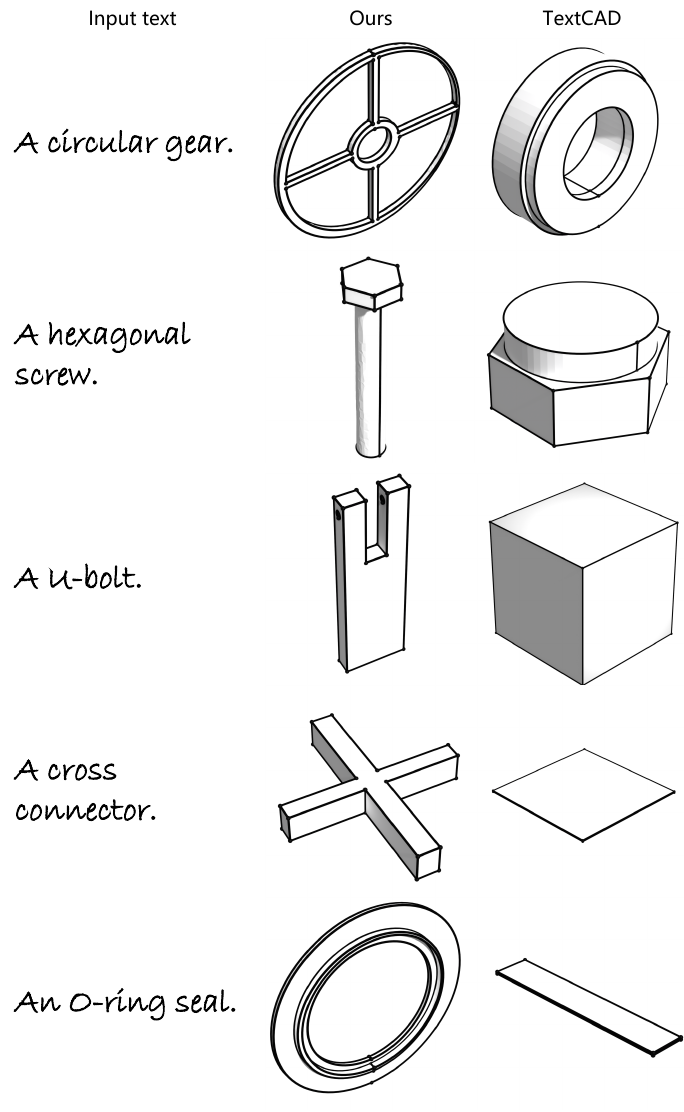}
    \caption{
        Qualitative results of text-based conditioned generation using \textbf{out-of-distribution} text description.
    }
    \label{fig:qualitative_od}
\end{figure}

\subsection{Text-conditioned B-Rep Generation}
\label{sec:results_text}
\paragraph{Dataset and Baselines.}
We evaluate our text-conditioned generation results on the Text2CAD dataset~\cite{khan2024textcad}, which builds upon the DeepCAD dataset by providing textual descriptions for each shape at four levels of complexity. Our model is trained on the Text2CAD training set until the validation loss converges. For comparison, we use the baseline method proposed in Text2CAD~\cite{khan2024textcad}, leveraging the official implementation and pre-trained models provided by the authors to generate B-Rep models from the text descriptions.

\paragraph{Discussion.}
As text descriptions provide only weak supervision and often contain ambiguities, we do not include quantitative results in this section. Instead, we present qualitative results in Fig.~\ref{fig:qualitative_txt}, demonstrating that our method generates more plausible models based on the text descriptions. Additionally, we evaluate the model's performance on out-of-distribution text descriptions in Fig.~\ref{fig:qualitative_od}. Even with more abstract and less detailed descriptions than those in the dataset, our method still generates plausible models, highlighting its robustness to varying levels of detail in the input text.

\paragraph{Failures.}
Fig.~\ref{fig:qualitative_txt_failed} shows six failure cases. These failures are primarily due to misalignment between the CAD models and the detailed descriptions, particularly regarding the number of components, symmetry constraints, and specific geometric details.

\subsection{Image-Conditioned B-Rep Generation}
\label{sec:results_images}

\paragraph{Dataset.}
For image-conditioned generation, we use three different types of inputs: single-view image, multi-view images, and sketches. As there is no publicly available dataset with rendered images for the DeepCAD dataset, we use the rendering engine in OpenCascade to generate images. For each model, we collect eight images from different views and one sketch image. Our model is trained on the rendered images until the validation loss converges. Since no open-source implementation for image-to-CAD generation is available, we provide qualitative results and a metric summary in this section.

\begin{figure*}
    \centering
    \includegraphics[width=1\linewidth]{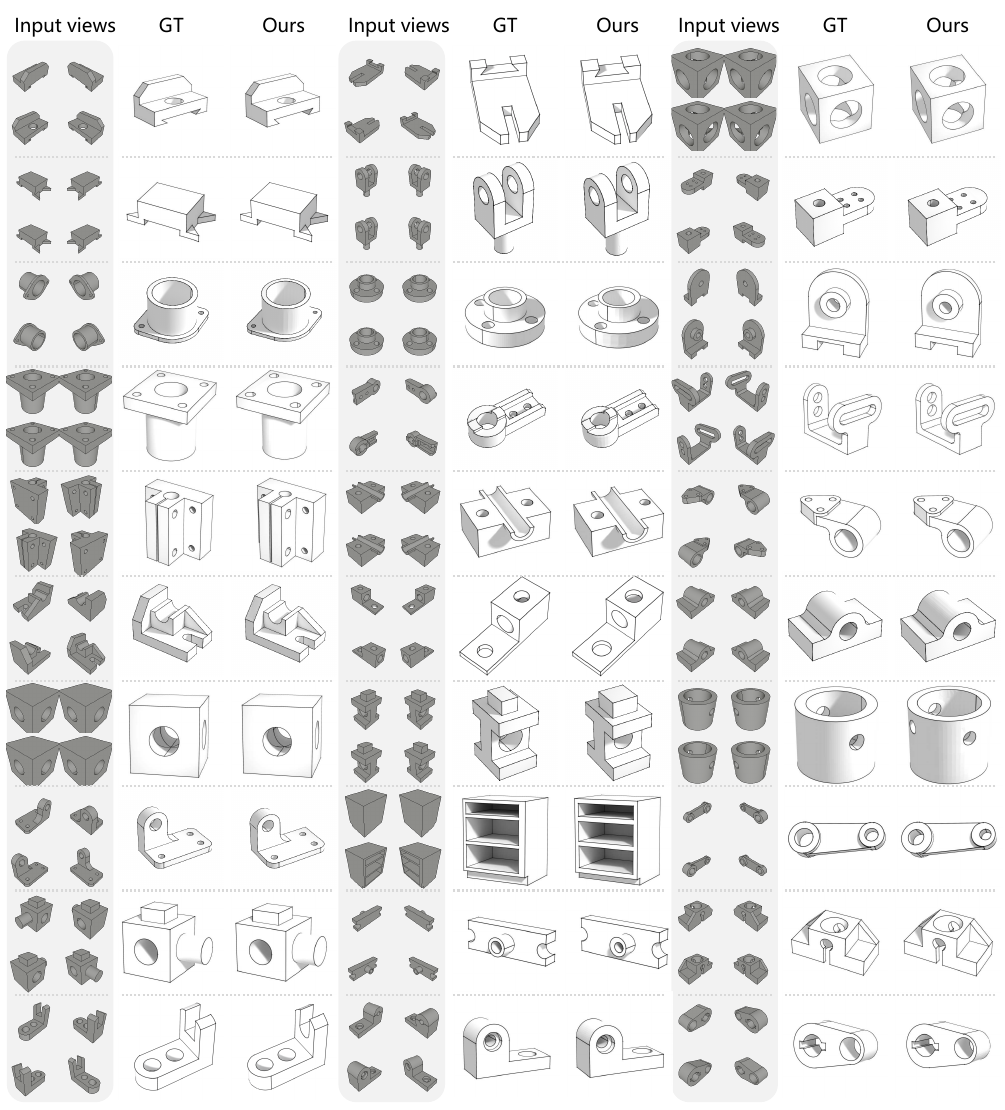}
    \caption{
        Qualitative results of multi-view conditioned generation.
    }
    \label{fig:qualitative_mvr}
\end{figure*}

\begin{figure*}
    \centering
    \includegraphics[width=1\linewidth]{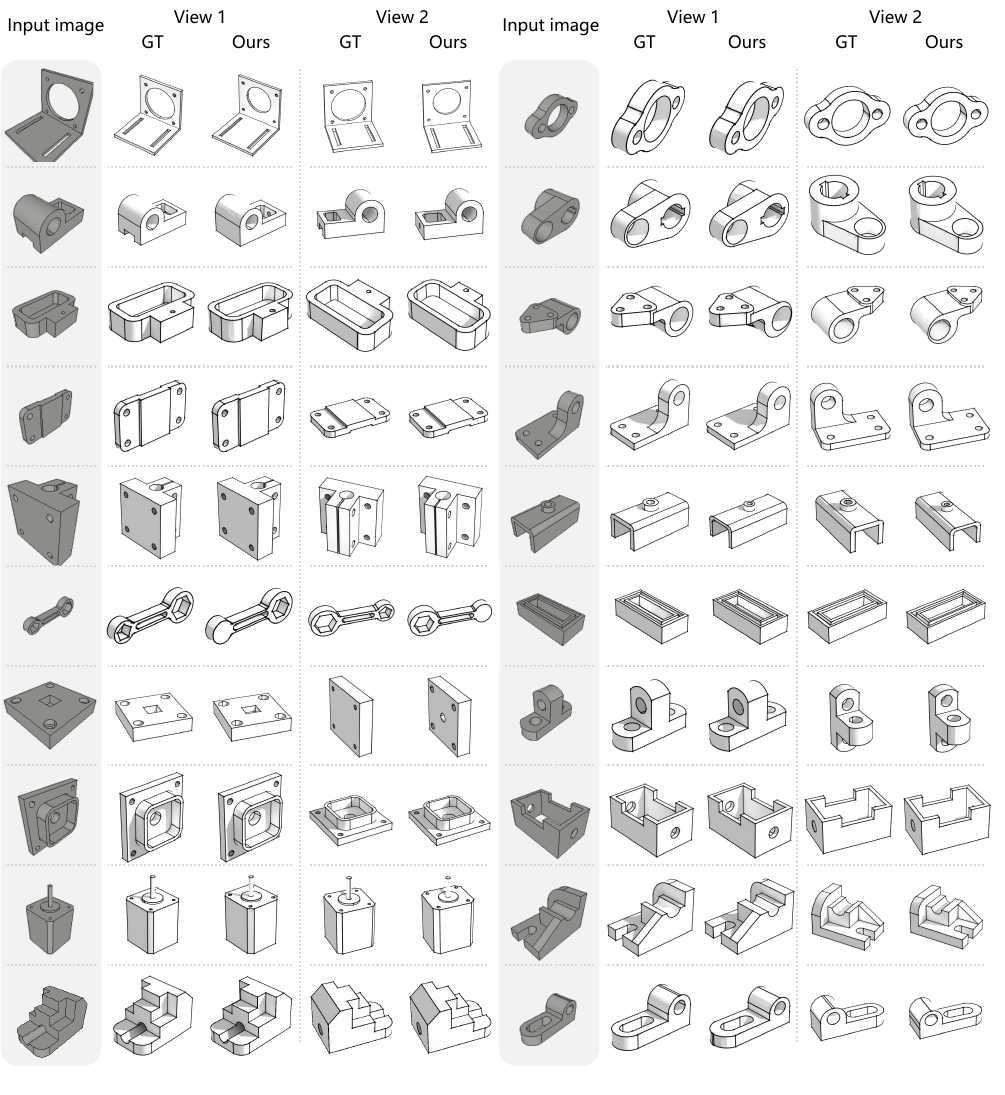}
    \caption{
        Qualitative results of single-view conditioned generation.
    }
    \label{fig:qualitative_svr}
\end{figure*}

\begin{figure*}
    \centering
    \includegraphics[width=1\linewidth]{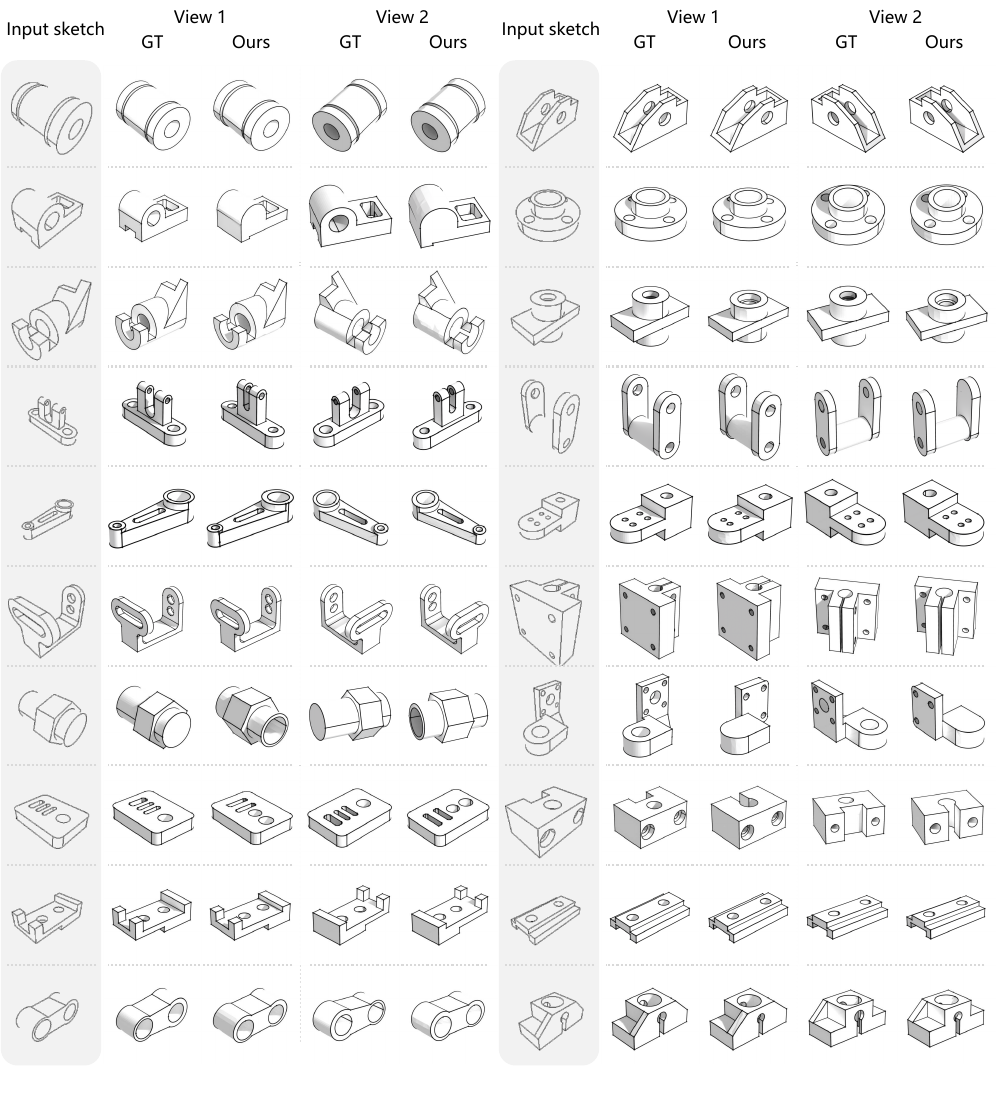}
    \caption{
        Qualitative results of sketch-based conditioned generation.
    }
    \label{fig:qualitative_sketch}
\end{figure*}

\paragraph{Discussion.}
Table~\ref{tab:quantitative_images} presents the quantitative results for image-conditioned generation. As expected, multi-view conditioned generation achieves the best performance in terms of Chamfer distance and F-scores for both geometry and topology. Qualitative results for multi-view, single-view, and sketch-based conditioned generation are shown in Fig.~\ref{fig:qualitative_mvr}, Fig.~\ref{fig:qualitative_svr}, and Fig.~\ref{fig:qualitative_sketch}, respectively.

\begin{figure*}
    \centering
    \includegraphics[width=0.95\linewidth]{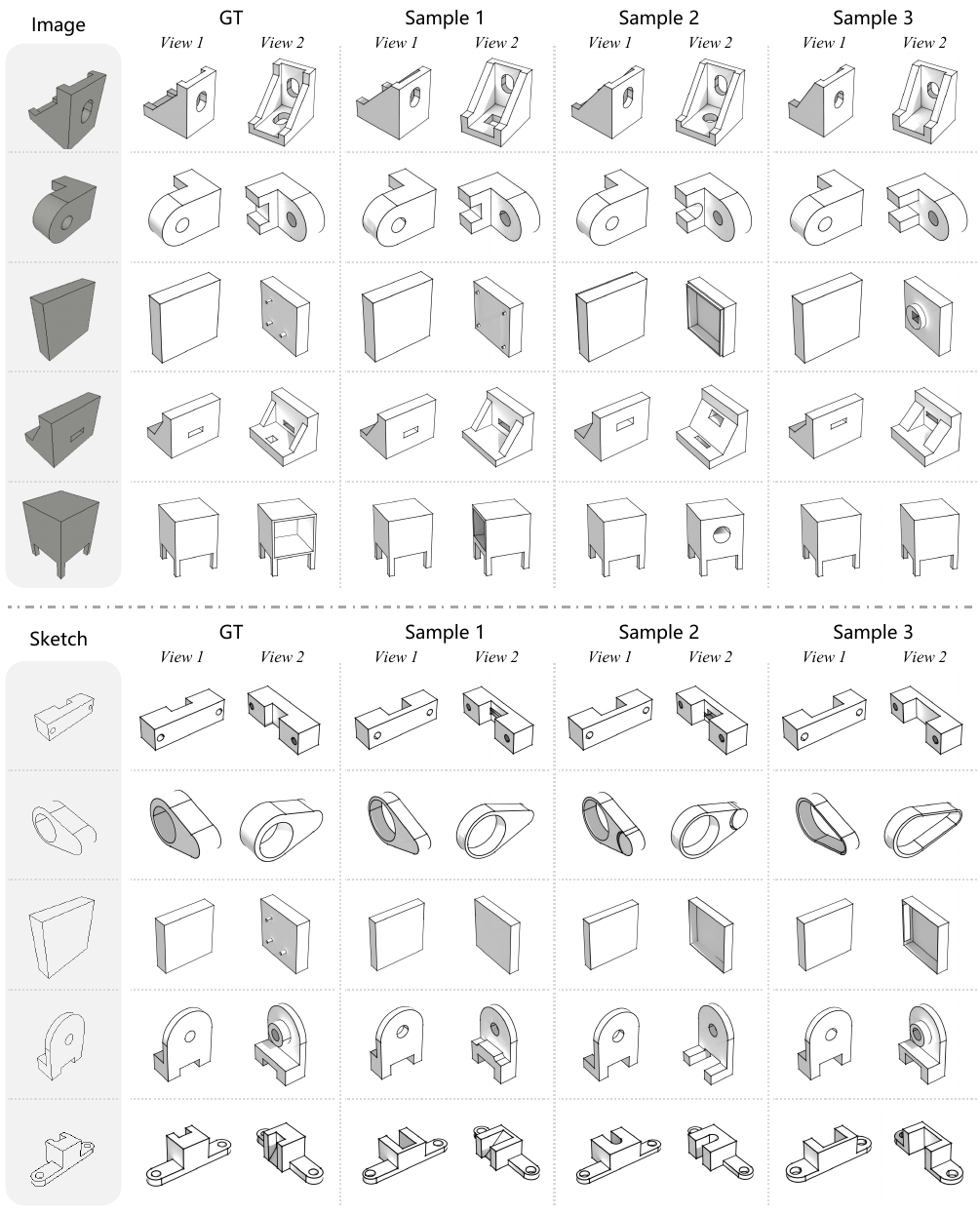}
    \caption{
        Multiple sampled generation results from the same single-view and sketch-based input.
    }
    \label{fig:qualitative_multigen}
\end{figure*}

Additionally, Fig.~\ref{fig:qualitative_multigen} showcases multiple sampled results from the same single-view and sketch-based inputs. The model generates multiple plausible candidates based on the visible parts of the input while introducing variations in the unseen regions, demonstrating its ability to capture uncertainty in invisible areas.
More results are provided in the supplementary material.

\paragraph{Failures.}
Failure cases for image-conditioned generation are shown in Fig.~\ref{fig:qualitative_failure}. Similar to text-conditioned generation, these failures primarily result from misalignment between the CAD model and the input image, as well as inconsistencies in the generated surface primitives.

\begin{table}
    \centering
    \resizebox{\columnwidth}{!}{
        \begin{tabular}{l|c|c|c|c|c|c|c|c|c}
            \toprule
            Method      & \multicolumn{3}{c|}{Chamfer ↓} & \multicolumn{3}{c|}{Geometry ↑} & \multicolumn{2}{c|}{Topology ↑} & \multirow{2}{*}{Val.}                                                                                    \\
            \cmidrule(r){2-4} \cmidrule(r){5-7} \cmidrule(r){8-9}
            Name        & Vertex                         & Edge                            & Face                            & Vertex                & Edge          & Face          & FE            & EV            &                  \\
            \midrule
            Single-view & 0.0684                         & 0.0182                          & 0.0101                          & 0.79                  & 0.82          & 0.86          & 0.81          & 0.8           & \textbf{86.37\%} \\
            Multi-view  & \textbf{0.0451}                & \textbf{0.0119}                 & \textbf{0.0057}                 & \textbf{0.82}         & \textbf{0.84} & \textbf{0.87} & \textbf{0.81} & \textbf{0.81} & 84.03\%          \\
            Sketch      & 0.0770                         & 0.0216                          & 0.0128                          & 0.77                  & 0.79          & 0.84          & 0.79          & 0.78          & 86.28\%          \\
            \bottomrule
        \end{tabular}
    }
    \caption{
            Quantitative results of image-conditioned generation.
    }
    \label{tab:quantitative_images}
\end{table}

\subsection{Ablation}

We conducted an ablation study on the DeepCAD test set to assess the contribution of key components in our holistic VAE, with results presented in Table~\ref{tab:quantitative_ablation}. 
We report reconstruction accuracy and completeness errors for both geometry and topology, excluding intersection accuracy as it remains consistently high (99.99\%).
Specifically, we compare the full model against three variants: (1) without self-attention after latent sampling, (2) without self-attention and spatial resolution in the latent vector, and (3) without self-attention, spatial resolution, and half-edge representation.

Removing self-attention results in a lighter model with faster training. Although this variant shows slight improvements in accuracy and completeness, it suffers from reduced precision and recall, leading to a lower validity ratio. Performance degrades significantly when spatial resolution and the half-edge representation are removed. This indicates that directional cues—such as edge and face orientation—are essential for learning B-Rep topology. Standard pooling and flattening operations in VAEs are permutation-invariant and thus discard such directional information. Without preserving spatial structure and explicitly modeling direction via half-edges, the model fails to capture the underlying topology of B-Rep models.

\subsection{Limitation}
\label{sec:results_limitation}
While our holistic latent representation effectively integrates different geometry categories and their topological connections into a unified latent space, the space itself is still defined by surface primitives. Inconsistencies may arise in the generated surface primitives, resulting in inaccurate trimming and, ultimately, non-watertight B-Rep models.

Furthermore, since the number of surface primitives is variable, both BRepGen and our method rely on padding the primitives to a fixed number. Although random padding has been shown to significantly outperform zero padding, the padding-and-deduplication strategy introduces noise and ambiguity during training, which can propagate to the generated surface primitives. This issue is evident in Figs.~\ref{fig:qualitative_unconditional}, \ref{fig:qualitative_failure}, and \ref{fig:qualitative_txt_failed}.

In conditional generation, the condition signal is encoded as a simple 1D feature vector, which may not fully capture the complexity or intrinsic features of the input. Consequently, the model may fail to leverage the condition signal effectively, leading to inconsistencies between the input and the generated B-Rep model.

\section{Conclusions}
\label{sec:conclusion}
We present HoLa, a novel representation learning method for modeling B-Rep models using a holistic latent space. By tying the topological connections between different primitives to the geometry of lower-order primitives, these relationships are seamlessly integrated into the reconstruction of the geometry. This fusion of geometry and topology within a unified latent space enables more accurate and expressive B-Rep generation. Additionally, the use of a single latent space simplifies the training process and enhances the quality and diversity of conditional B-Rep generation.

In future work, we aim to explore compressing the latent space further into a global latent vector representing the entire B-Rep model. This approach could yield a more compact representation while potentially improving the quality and validity of the generated B-Rep models. Furthermore, incorporating a more advanced feature injection mechanism during conditional generative model training may enhance the quality of generated B-Rep models across various input conditions.

\section*{Acknowledgments}
We thank all the anonymous reviewers for their insightful comments. Thanks also go to Jinfeng Ou and Jiacheng Ren for the illustrations and Dani Lischinski, Xueqi Ma and Shanshan Pan for the helpful discussion.
This work was supported in parts by ICFCRT (W2441020), NSFC (U21B2023), Guangdong Research Foundation (2023B1515120026), Shenzhen Science and Technology Program (KJZD20240903100022028, KQTD20210811090044003, \seqsplit{RCJC20200714114435012}), and Scientific Development Funds from Shenzhen University.

\bibliographystyle{ACM-Reference-Format}
\bibliography{HoLaBrep}

\clearpage
\appendix
\section{Supplementary Material}
\begin{figure*}[!hbp]
    \centering
    \includegraphics[width=0.95\linewidth]{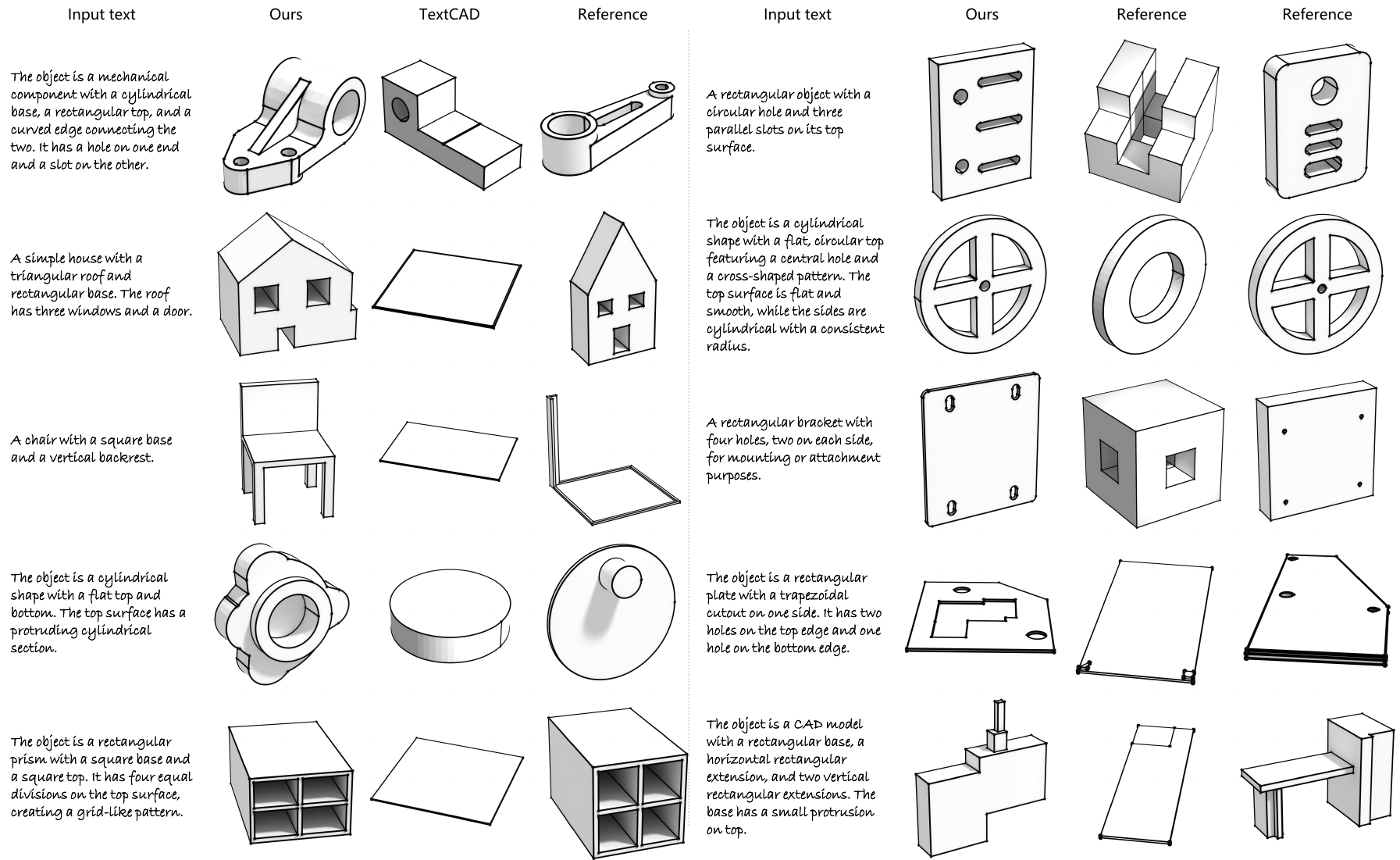}
    \caption{
        Additional results of text-conditioned generation.
    }
    \label{fig:cond_qualitative_txt_supp1}
\end{figure*}

\begin{figure*}[!hbp]
    \centering
    \includegraphics[width=1\linewidth]{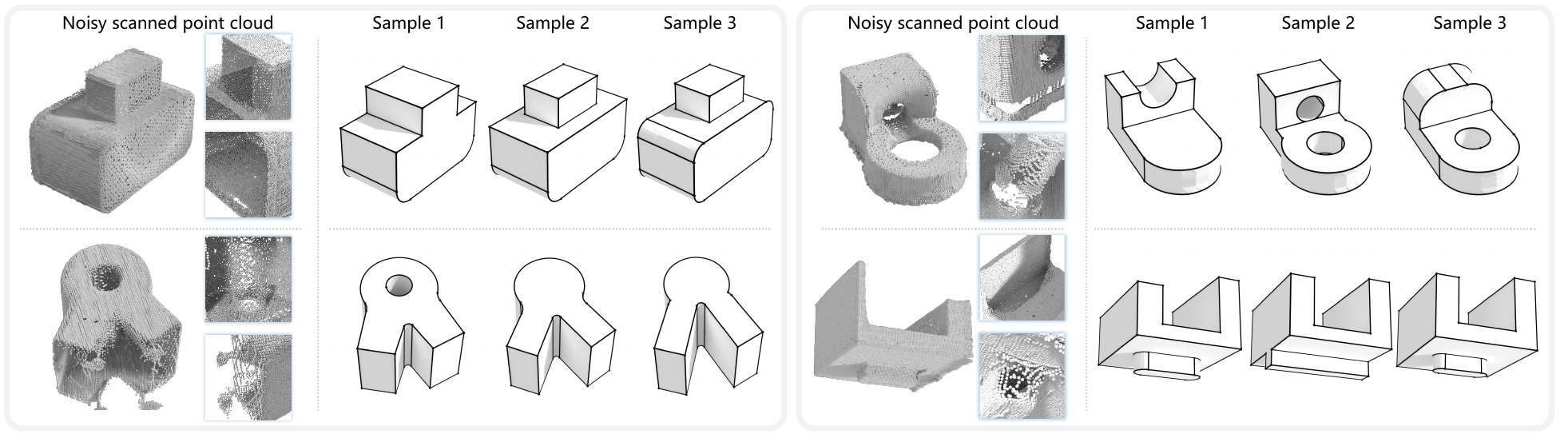}
    \caption{
        Additional results of point-conditioned generation.
    }
    \label{fig:qualitative_pc_real2}
\end{figure*}

\begin{figure*}[!hbp]
    \centering
    \includegraphics[width=0.9\linewidth]{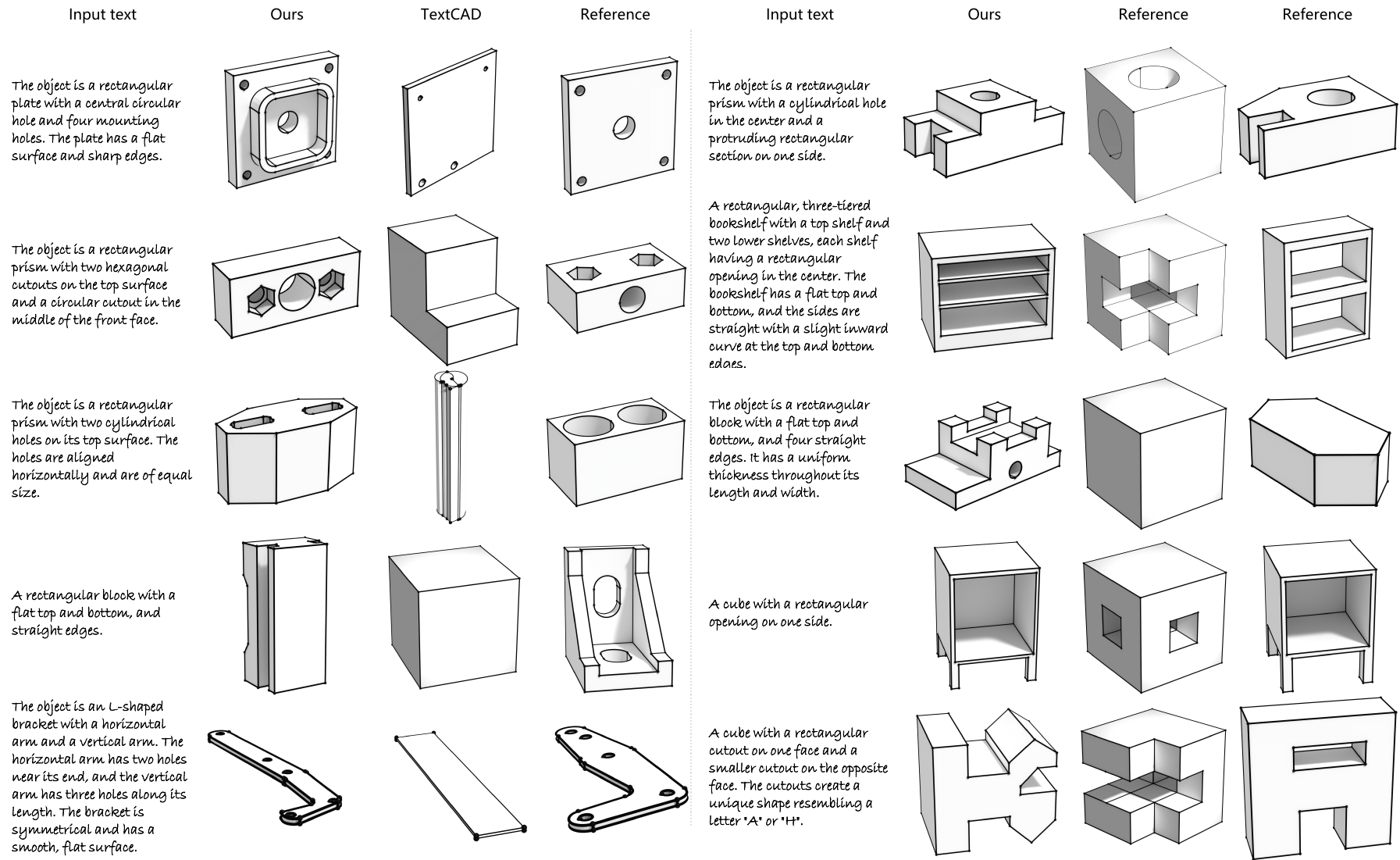}
    \caption{
        Additional results of text-conditioned generation.
    }
    \label{fig:cond_qualitative_txt_supp2}
\end{figure*}

\begin{figure*}[!hbp]
    \centering
    \includegraphics[width=0.9\linewidth]{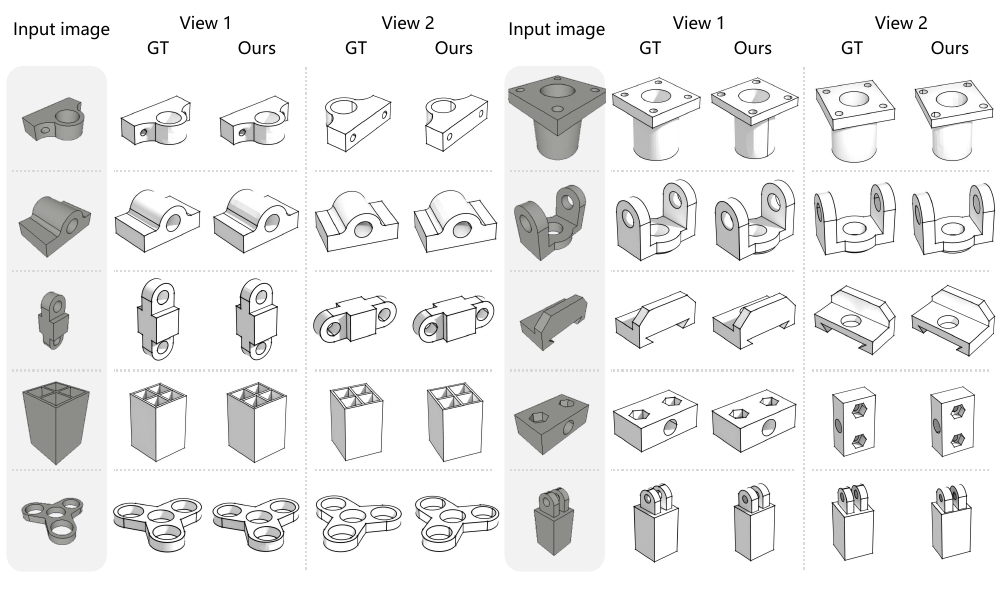}
    \caption{
        Additional results of single-view conditioned generation.
    }
    \label{fig:cond_qualitative_svr_supp2}
\end{figure*}

\begin{figure*}[!hbp]
    \centering
    \includegraphics[width=0.9\linewidth]{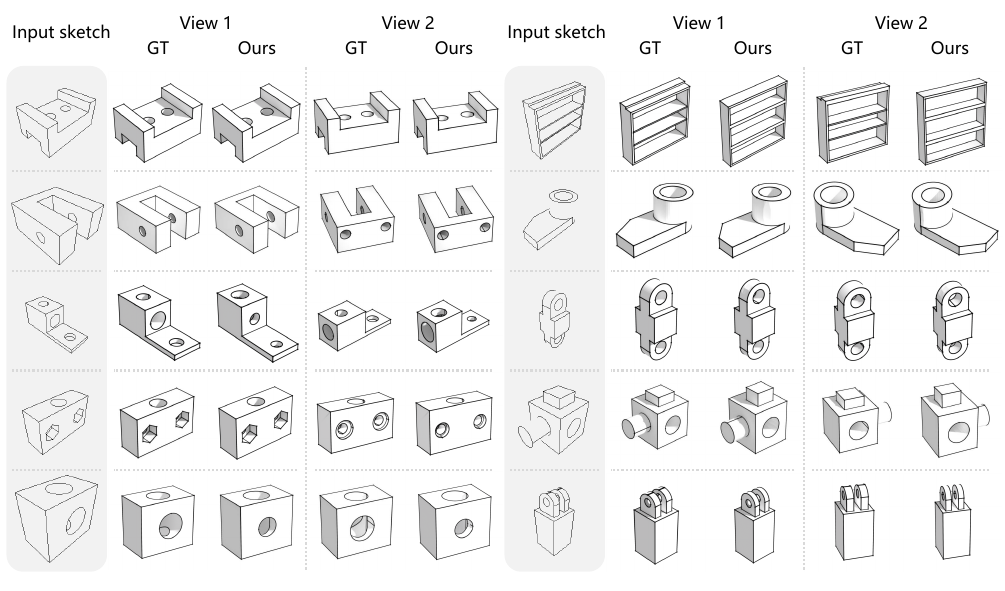}
    \caption{
        Additional results of sketch-conditioned generation.
    }
    \label{fig:cond_qualitative_sketch_supp}
\end{figure*}

This supplementary material presents additional results and analysis for our proposed holistic latent representation (HoLa).
We show additional point-conditioned generation results from the real-world scanned data in Fig.~\ref{fig:qualitative_pc_real2}, text-conditioned generation results in Fig.~\ref{fig:cond_qualitative_txt_supp1} and Fig.~\ref{fig:cond_qualitative_txt_supp2}, single-view and sketch-conditioned generation results in Fig.~\ref{fig:cond_qualitative_svr_supp2}, and Fig.~\ref{fig:cond_qualitative_sketch_supp}, respectively.

\end{document}